\documentclass[runningheads,a4paper,12pts]{llncs}

\usepackage{amssymb}
\usepackage{graphicx}
\usepackage{footnote}
\makesavenoteenv{tabular}
\makesavenoteenv{table}
\usepackage{url}

\begin{document}

\mainmatter  

\title{Simulating radially outward winds within a turbulent gas clump}

\titlerunning{winds within clump}

\author{Guillermo Arreaga-Garc\'{\i}a$^{1}$ \and Silvio Oreste Topa$^{2}$ }
%
\authorrunning{Arreaga-Garc\'{\i}a and Topa}

\institute{Departamento de Investigaci\'on en F\'{\i}sica\\
Universidad de Sonora\\
Hermosillo, Sonora, MEXICO\\
\email{garreaga@cifus.uson.mx}
\and
EESA Num. 1, \\
Tornquist, Pcia. de Buenos Aires\\
ARGENTINA }

\toctitle{Simulating winds within a clump}
\tocauthor{Arreaga-Garcia, Topa}
\maketitle

\begin{abstract}
By using the particle-based code
Gadget2, we follow the evolution
of a gas clump, in which a gravitational collapse is initially induced.
The particles representing the gas clump have initially
a velocity according to a turbulent spectrum built in a
Fourier space of 64$^3$ grid elements. In a very early stage of
evolution of the clump, a set of gas particles representing
the wind, suddenly move outwards from the clump's center. We consider
only two kinds of winds, namely: one with spherical symmetry and a
second one being a bipolar collimated jet. In order to assess the dynamical
change in the clump due to interaction with the winds, we show
iso-velocity and iso-density plots for all our simulations.
\end{abstract}

\section{Introduction}

Stars are born in large gas structures made of molecular hydrogen.
These gas structures are named clumps; see Ref.~\cite{bergin}. These clumps 
have typical sizes and masses of a few pc (parsecs) and a few 
hundred or even thousands of $M_{\odot}$ (one solar mass), 
respectively\footnote{A parsec (pc) is equivalent to $3.08 \, \times
10^{18}\, $cm and a solar mass $M_{\odot}$ is equivalent to $1.99 \,
\times 10^{33}\, $g.}.

The physical process by which the molecular gas is transformed from
a gas structure into some stars is mainly gravitational collapse, whose main effects
on the gas clump are that the gas density is increased while its size is reduced.
At some point during this transformation process from gas to
star, the densest gas structures settled down in more or less dynamically
stable gas objects called protostars\footnote{Any star does radiate its own
energy produced by thermonuclear reactions in its interior, but a
protostar does not. This is the main difference between a star and a
protostar; but they can share some dynamical properties as they are
stable structures of different stages of the same formation
process.}. For instance, the number density of a typical large clump
structure ranges around $10^3$ molecules per cm$^{-3}$ whilst that
of a typical protostar ranges around $10^{14}$ molecules per
cm$^{-3}$. To achieve a better understanding of the huge change in
scales, we mention that the number density of a typical star is
around $10^{24}$ molecules per cm$^{-3}$. The results of a set of numerical 
simulations aimed to study the gravitational collapse of a
spherically symmetric, rigidly rotating, isolated, interstellar gas core 
are presented in Ref.~\cite{jaime}. 

The process of gravitational collapse is not the only process acting upon the gas
structures in the interstellar medium, as many other phenomena can
have a great influence on the evolution of the clump, among others: (i) the highly ionized gas
ejected by the explosion of supernovas; (ii) the bipolar collimated
winds ejected by a massive protostar; (iii) rapidly expanding
H$_{II}$ regions which hit the slower and less dense gas structures.

Recently, in Ref.\cite{dale}, a set of SPH simulations were
conducted to study star formation triggered by an expanding H$_{II}$
region within a spherical gas structure with uniform density that
was not globally collapsing. The expanding shock plays the role of a
snowplow, which sweeps out the gas of the surrounding medium. The
density of the swept out gas increases as a consequence of this
agglomeration process, and a gravitational collapse may then be
locally initiated. Small gas over-densities can be formed in this
way, which may achieve the protostar stage. Because these two
processes are complementary in forming proto-stars, this star
formation scenario was named the collect and collapse model.

Furthermore, in order to study the effects of proto-stellar outflows
on the turbulence of a star forming region, in Refs.\cite{naka06}
and \cite{naka07}, magneto-hydrodynamics simulations (MHD) were
conducted with a mesh based code which implements the Adaptive Mesh
Refinement technique (AMR).

In this paper, we investigate the change in the dynamical configuration of a typical
turbulent clump when a wind of particles is outwardly ejected from the
central region of the clump. The most important difference
between the present paper and Ref. \cite{dale} is
the turbulent nature of our clump. Turbulence makes a big change in the
spatial distribution as the clump becomes
filamentary and flocculent.

\section{The physical system}
\label{sec:physys}

In this section we briefly describe the physics of the clump and the
winds, which will be considered in the following sections.
\subsection{The initial clump}
\label{sec:theclump}

We here consider a typical spherical clump with a radius $R_0=2\, $
pc and mass $M_0=1219\, M_{\odot}$. Initially, it has a radially
uniform density distribution with an average density given by
$\rho_0=2.4 \times 10^{-21} \; $ g $\,$ cm$^{-3}$, which is
equivalent to a number density $n_0 \approx 600\,$ molecules
cm$^{-3}$ for molecular hydrogen with molecular mass $\mu=4.0\times
10^{-24}\,$gr/mol. The size and mass of this clump are chosen here
to be typical in the statistical sense, in accordance with
Ref.~\cite{bergin}.

The free fall time  $t_{ff}$ is defined as the time needed for an external particle to reach
the center of the clump when gravity is the only force pulling the particle. In this idealized
gravitational collapse, we have

\begin{equation}
t_{ff} \approx \sqrt{ \frac{3\, \pi}{32 \, G \, \rho_0}}
\label{tff}
\end{equation}
\noindent where $G$ is Newton's universal gravitational constant. For our clump, we
have $t_{ff}=1.3 \times 10^6\,$ years.

Following Ref. \cite{bodeniv}, the dynamical properties of the
initial distribution of the gas are usually characterized by $\alpha$, the
ratio of the thermal energy to the gravitational energy, and $\beta$, that of the rotational
energy to the gravitational energy. For a spherical clump, the approximate
total gravitational
potential energy is $<E_{grav}> \approx - \frac{3}{5} \; \frac{G\, M_0^2}{R_0}$. The
average total thermal
energy $<E_{therm}>$ ( kinetic plus potential interaction terms of the molecules) is
$ <E_{therm}> \approx \frac{3}{2} {\cal N} \, k \, T = \frac{3}{2} M_0\, c_0^2$,
where $k$ is the Boltzmann constant, $T$ is the equilibrium temperature,
${\cal N}$ is the total number of molecules in the gas and $c_0$ is the
speed of sound, see Section ~\ref{subs:energies} for a more precise definition.

The kinetic energy $<E_{kin}>$ can be estimated by $M_0\, v_{av}^2/2$, where $v_{av}$ is
the average translational velocity of the clump. In order to
have both energies of the same order of magnitude, $ <E_{kin}> \approx <E_{grav}> $, the
gas elements of the clump must attain average velocities within the range

\begin{equation}
v_{av}/c_0 \approx 3-3.5
\label{vavrange}
\end{equation}
\noindent or $v_{av} \approx 1.6 \,$km/s, for a speed of sound given
by

\begin{equation}
\begin{array}{l}
c_0= 0.54 \, {\rm km}/{\rm s} \; \equiv 54862.91 \; {\rm cm}/{\rm s}
\end{array}
\label{sound}
\end{equation}
\noindent so that the corresponding temperature associated with the clump
is $T \approx 25 \,$K.

It is possible to define the crossing time by means of

\begin{equation}
t_{cr} \approx \frac{R_0}{c_0}=3.56\, \times 10^6 \, {\rm yr}\,
\label{tcr}
\end{equation}
\noindent which sets a time scale for a sound wave to travel across
the clump. To make the crossing time comparable in magnitude to the
free fall time of Eq.~\ref{tff}, the front wave must have velocities
around $v_{req}/c_0 \approx 2.6 $ or $v_{req} \approx 1.45 \, $
km/s, which are velocities a little bit slower than the ones
estimated above, see Eq.~\ref{vavrange}. Anyway, in this paper we
will treat propagation velocities of gas particles ranging around
$2-3$ Mach.

\subsection{The wind.}
\label{subsec:winds}

In this paper we consider two kinds of winds: the first kind has a fully spherical
symmetry and the second kind is a bipolar collimated jet.

The dynamical characteristics of the wind strongly depends on its
type of source. All stars eject winds of the first kind, which are
driven by the stellar radiation. For instance, in cool stars, like
the ones observed in the AGB (asymtotic giant branch) of the Galaxy,
the winds cause a mass loss in the range $10^{-8}-10^{-4}\,
M_{\odot}/$yr whereas the terminal wind velocities are around
$10-45\, $km/s. In OB stars, the mass loss ranges over  $10^{-6}-
10^{-4} \, M_{\odot}$/yr and the terminal wind velocities can go up
to thousands of km/sec.

Supernovas dump around $10^4$ joules of thermal and kinetic energy
into the interstellar medium. But there are many types of
supernovas, so that the mass losses and terminal velocities are very
different, see Ref. \cite{supernovas}. For example, for a supernova
whose progenitor was a He star, the mass loss and terminal
velocities are within the ranges  $10^{-7} - 10^{-4} \,
M_{\odot}$/yr and $100-1000$ km/sec, respectively. When the
progenitor was a RSG star, then their values ranges over $10^{-5} -
10^{-4} \, M_{\odot}$/yr and $10-40$ km/s, respectively.

It seems that all protostars eject highly collimated jets of gas
during their formation process by gravitational collapse. The origin
of these jets is still unclear but it may be that the accretion disk
and magnetic field around the protostars play a crucial role in
determining the velocities and the degree of collimation of the jets.
For the molecular winds associated with protostarts of Class 0
and Class 1, the characteristic velocities are around 20 km/sec. However, for optical
jets of highly ionized gas, the typical jet velocities are a few hundred km/sec. See
Ref.~\cite{jet} and the references therein.

\section{The Computational Method}
\label{subsec:met}

In this section we briefly describe the way we set up the physical system outlined above
in computational terms.
\subsection{The initial configuration of particles.}
\label{subsec:inicond}

We set $N=10$ million SPH particles for representing the
gas clump. By means of a rectangular
mesh we make the partition of the simulation volume in small
elements each with a volume $\Delta x\, \Delta y\, \Delta z $; at
the center of each volume we place a particle (the $i$th, say), with
a mass determined by its location according to the density profile
being considered, that is: $m_i= \rho(x_i,y_i,z_i)*\Delta x\, \Delta
y\, \Delta z$ with $i=1,...,N$.  Next, we displace each particle from
its location by a distance on the order of $\Delta x/4.0$ in a random
spatial direction.

As was stated earlier, in this paper we only consider a uniform density clump, for
which $\rho(x_i,y_i,z_i) \equiv \rho_0$, for all the simulations (see
Section~\ref{sec:theclump}). Therefore, all the particles have the same mass
irrespective of whether a wind or clump particle.

\subsection{The initial turbulent velocity of particles.}
\label{subsec:turbspect}

To generate the turbulent velocity spectrum for the clump particles, we follow
a procedure based on the papers \cite{dubinski} and \cite{dobbs}. We set
a second mesh $N_x,N_y,N_z$ with the size of each element given by $\delta x=R_0/N_x$, $\delta
y=R_0/N_y$ and  $\delta z=R_0/N_z$. In Fourier space, the partition
is $\delta K_x=1.0/\left( N_x \times \delta x \right)$ , $\delta
K_y=1.0/\left( N_y \times \delta y \right)$ and $\delta
K_z=1.0/\left( N_z \times \delta z \right)$. Each Fourier mode has
the components $K_x=i_{K_x}  \delta K_x$, $K_y=i_{K_y} \, \delta K_y
$ and $ K_z=i_{K_z} \, \delta K_z $, where the indices $i_{K_x}$,
$i_{K_y}$ e $i_{K_z}$ take values in $[-N_x/2,N_x/2]$,
$[-N_y/2,N_y/2]$ and $[-N_z/2,N_z/2]$, respectively. The wave number
magnitude is $K=\sqrt{K_x^2+K_y^2+K_z^2}$, and so $K_{min}=0 $ and
$K_{max}= \frac{\sqrt{3}\, N_x}{2 \, R_0} $. The Fourier wave can
equally be described by a wave length $\lambda=2\, \pi/K$, then we
see that $K \approx \frac{1}{R_0}$ and $\lambda \approx R_0$.

Following \cite{dobbs}, the components of the particle velocity are

\begin{equation}
\begin{array}{l}

\vec{v}=\Sigma_{-K_{max}}^{K_{max}} \left| K^{ \frac{-n-2}{2} }\right|
\times

\left\{

\begin{array}{l}

\left[ K_z \, C_{K_y} \sin \left( \vec{K}\cdot \vec{r} + \Phi_{K_y}\right) -
K_y \, C_{K_z} \sin \left( \vec{K}\cdot \vec{r} + \Phi_{K_z}\right)\right] \mbox{for}\, v_x\\

\left[ - K_x \, C_{K_z} \sin \left( \vec{K}\cdot \vec{r} + \Phi_{K_z}\right) +
K_z \, C_{K_x} \sin \left( \vec{K}\cdot \vec{r} + \Phi_{K_x}\right)\right] \mbox{for} \, v_y\\

\left[ -K_x \, C_{K_y} \sin \left( \vec{K}\cdot \vec{r} + \Phi_{K_y}\right) +
K_y \, C_{K_x} \sin \left( \vec{K}\cdot \vec{r} + \Phi_{K_x}\right)\right] \mbox{for} \, v_z\\
\end{array}

\right.
\end{array}
\label{velturb}
\end{equation}

\noindent where the spectral index $n$ was fixed at $n=-1$ and thus
we have $v^2\approx K^{-3}$.  The vector $\vec{C}_{K}$ whose components
are denoted by $\left( C_{K_x}, C_{K_y}, C_{K_z}\right)$, take
values obeying a Rayleigh distribution. The wave phase vector, $\Phi_K$,
given by $\left( \Phi_{K_x}, \Phi_{K_y}, \Phi_{K_z} \right)$ takes
random values on the interval $[0,2\,\pi]$. The components of the vector $\vec{C}$
are calculated by means of  $C=\sigma\times \sqrt{ -2.0
\times \log \left(1.0-u\right) }$, where $u$ is a random number in
$(0,1)$. $\sigma$ is a fixed parameter with value $\sigma=1.0$.

\subsection{The set up of the particle wind.}
\label{subsec:setupwinds}

Let us consider the equation of mass conservation for a set of particles moving radially
outwards, that is

\begin{equation}
\dot{M}=4\, \pi \, r^2 \, \rho(r)\times v(r)
\label{consmasa}
\end{equation}
\noindent  We fix the mass loss $\dot{M}$ as a parameter of the simulation and also fix the
wind density to have the uniform value $\rho_0$. We then determine the wind
velocities according to Eq.~\ref{consmasa}. As the velocity magnitude diverges for
particles around $r \approx 0$, we set a cut velocity value such that the maximum
velocity allowed in our simulations is $v_{max}$.

Of course, there are other possibilities, which will be considered
elsewhere: one is to fix the radial density $\rho(r)$ and/or the
velocity profile $v(r)$ in order to obtain the mass loss $\dot{M}$
as a result. Besides, for modeling an expanding $H_{II}$ region, the
authors of Ref. \cite{dale} proposed another and more complicated
velocity function, but anyway it gives a constant expansion velocity
at the last stages of time evolution, so that the average velocity
of the shocked shell considered by \cite{dale} is $v_{cc}/c_0
\approx 5.6 $ or $v_{cc} \approx 3.7 \, $ km/s.

\subsection{Initial energies}
\label{subs:energies}

In a particle based code, we approximate the thermal energy of the clump by calculating the sum
over all the $N$ particles described in Section~\ref{subsec:inicond}, that is

\begin{equation}
E_{therm}=\sum_{i=1}^{N} \, \frac{3}{2} \,
\frac{P_i(\rho)\,m_i}{\rho_i} \, ,
\end{equation}
\noindent where $P_i$ is the pressure associated with particle $i$ with density
$\rho_i$ by means of the equation of state given in Eq. \ref{beos}.
In a similar way, the approximate potential energy is

\begin{equation}
E_{pot}= \sum_{i=1}^{N}\, \frac{1}{2}\, m_i \, \Phi_i \, .
\end{equation}
\noindent where $\Phi_i $ is the gravitational potential of particle $i$. For the clump considered
in this paper, the values of the speed
of sound $c_0$ (see Eq.~\ref{sound} ) and the level of turbulence are chosen so
that the energy ratios have the numerical values

\begin{equation}
\begin{array}{l}
\alpha \equiv \frac{E_{therm}}{\left|E_{grav}\right|}=0.3 \vspace{0.25 cm}\\
\beta \equiv \frac{E_{kin}}{\left|E_{grav}\right|}=1.0
\end{array}
\label{alphaybeta}
\end{equation}
\subsection{Resolution and thermodynamical considerations}
\label{subs:resol}

Following Refs. \cite{truelove} and \cite{bateburkert97},
in order to avoid artificial fragmentation,
the $SPH$ code must fulfil certain resolution criteria, imposed on the Jeans
wavelength $\lambda_J$, which is given by

\begin{equation}
\lambda_J=\sqrt{ \frac{\pi \, c^2}{G\, \rho}} \; , \label{ljeans}
\end{equation}
\noindent where $c$ is the instantaneous speed of sound and $\rho$ is the
local density. To obtain a more useful form for a particle based code, the Jeans
wavelength $\lambda_J$ is transformed into the Jeans mass given by

\begin{equation}
M_J \equiv \frac{4}{3}\pi \; \rho \left(\frac{ \lambda_J}{2}
\right)^3 = \frac{ \pi^\frac{5}{2} }{6} \frac{c^3}{ \sqrt{G^3 \,
\rho} } \;. \label{mjeans}
\end{equation}
\noindent In this paper, the values of the density and speed of sound are updated
according to the following equation of state

\begin{equation}
p= c_0^2 \, \rho \left[ 1 + \left(
\frac{\rho}{\rho_{crit}}\right)^{\gamma -1 } \, \right] ,
\label{beos}
\end{equation}
\noindent as proposed by \cite{boss2000}, where $\gamma\, \equiv
5/3$ and for the critical density we assume the value
$\rho_{crit}=5.0 \times 10^{-14} \, $ g $\,$ cm$^{-3}$.

For the turbulent clump under consideration, we have $m_r \approx M_J
/ (2 N_{neigh}) \approx 7.47 \times 10^{33}\,$ g, where we take $N_{neigh}=40$.

In this paper, the mass of an SPH particle is $m_p=1.98\, \times 10^{29}\,$g, so
that $m_p/m_r=2.5\, \times 10^{-4}$ and therefore the Jeans
resolution requirement is satisfied very easily.

In a previous paper of collapse reported in Ref.\cite{NuestroApJ}, by means of a
convergence study, we demonstrated the correctness of a regular cartesian grid to
make collapse calculation, as is used here to make the partition of the simulation
domain in small volume elements, each of which has a SPH particle located not
necessarily in its center, see Sect.\ref{subsec:inicond}. 
\section{The evolution code}
\label{sec:thecode}

We carry out the time evolution of the initial distribution of
particles with the fully parallel Gadget2 code, which is described
in detail by Ref.\cite{gadget2}. Gadget2 is based on the $tree-PM$
method for computing the gravitational forces and on the standard
$SPH$ method for solving the Euler equations of hydrodynamics.
$Gadget2$ incorporates the following standard features: (i) each
particle $i$ has its own smoothing length $h_i$; (ii) the particles
are also allowed to have individual gravitational softening lengths
$\epsilon_i$, whose values are adjusted such that for every time
step $\epsilon_i \, h_i$ is of order unity. Gadget2 fixes the
value of $\epsilon_i$ for each time-step using the minimum value of
the smoothing length of all particles, that is, if
$h_{min}=min(h_i)$ for $i=1,2...N$, then $\epsilon_i=h_{min}$.

The $Gadget2$ code has an implementation of a Monaghan-Balsara form
for the artificial viscosity, see Ref.\cite{mona1983} and
Ref.\cite{balsara1995}. The strength of the viscosity is regulated
by setting the parameter $\alpha_{\nu} = 0.75$ and
$\beta_{\nu}=\frac{3}{2}\, \times \alpha_{\nu}$, see Equation (14)
in Ref.\cite{gadget2}. We here fix the Courant factor to 0.1.

We now mention here that the public Gadget2 code used in this
paper, presents some potential technical problems when wind particles
are simultaneously evolved with clump particles; these  problems are
caused by the different mass scales involved, as the discretized
version of the Navier-Stokes hydrodynamical equations are written
in the so called density-entropy formulation, see
Ref.\cite{hopkins}: mainly that the particle time-step becomes
prohibitively small to achieve the overall evolution of the system.

We finally mention that the initial condition code was
written in ANSI-C and it makes use of a Fourier mesh of 643 grid
elements in order to obtain the turbulent velocity field of 10
million SPH particles. It takes 300 CPU hours running on one
INTEL-Xeon processor at 3.2 GHz. All the simulations presented in
this work were conducted on a high performance Linux cluster with 46
dual cores located at the University of Sonora (ACARUS). The
simulations were calculated over 300 CPU hours on a fully
parallelized scheme using 12 cores per each run. The proprietary
program pvwave-8.0 was used as the main the visualization tool for
showing the results of this paper.
\section{Results}
\label{sec:res}

To present the results of our simulations, we consider a slice of
particles around the equatorial plane of the clump; with these
particles (around 10,000) we make plots containing two complementary
panels: one to show colored regions of iso-density and the other one
to show the velocity field of the particles. We also make $3D$ plots
built with the 500,000 densest particles of each simulation. Later,
we present plots with the velocity distributions and radial velocity
profile, for which we use all the simulation particles.

In Table ~\ref{tab:models} we show the values we use to define the
simulations and also we give some results described below. In this Table and in
all the subsequent figures, we use the following labels: "Tur" to
indicate the gas clump; when the winds appear within the clump, we
use the label "Tur+Wind" to refer to the spherically symmetric case
and the label "Tur+Wind+Col" for the bipolar jet.
\begin{table}[h]
\begin{center}
\caption{The models.}
\label{tab:models}
\begin{tabular}{|c|c|c|c|c|c|c|c|}
\hline \vspace{0.1 cm}
Label          & $t_{start}/t_{ff} $&  $\dot{M}\; [M_{\odot}/{\rm yr}]$ &  $\frac{R_s}{R_0}$ &  $M_g \,[M_{\odot}]$ & $M_{\infty} \,[M_{\odot}] $ &  $v_{max}/c_s$ & $v_{\infty}/c_s$\\
\hline
 T  & 0.0 &  ----- &-----& 1219 &  333 & 100 & 10  \\
\hline
  Tur+Wind     & 0.05 &  $1.0\, \times 10^{-4}$  &  $ 0.1 $  & 6.84  & 359 & 100   & 11\\
\hline
  Tur+Wind+Col & 0.05 &  $1.0\, \times 10^{-3}$  &  $ 0.05$  & 4.51  & 422  & 100 & 72 \\
\hline
\end{tabular}
\end{center}
\end{table}

\begin{figure}
\begin{center}
\begin{tabular}{cc}
\includegraphics[width=2.0in, height=2.0 in]{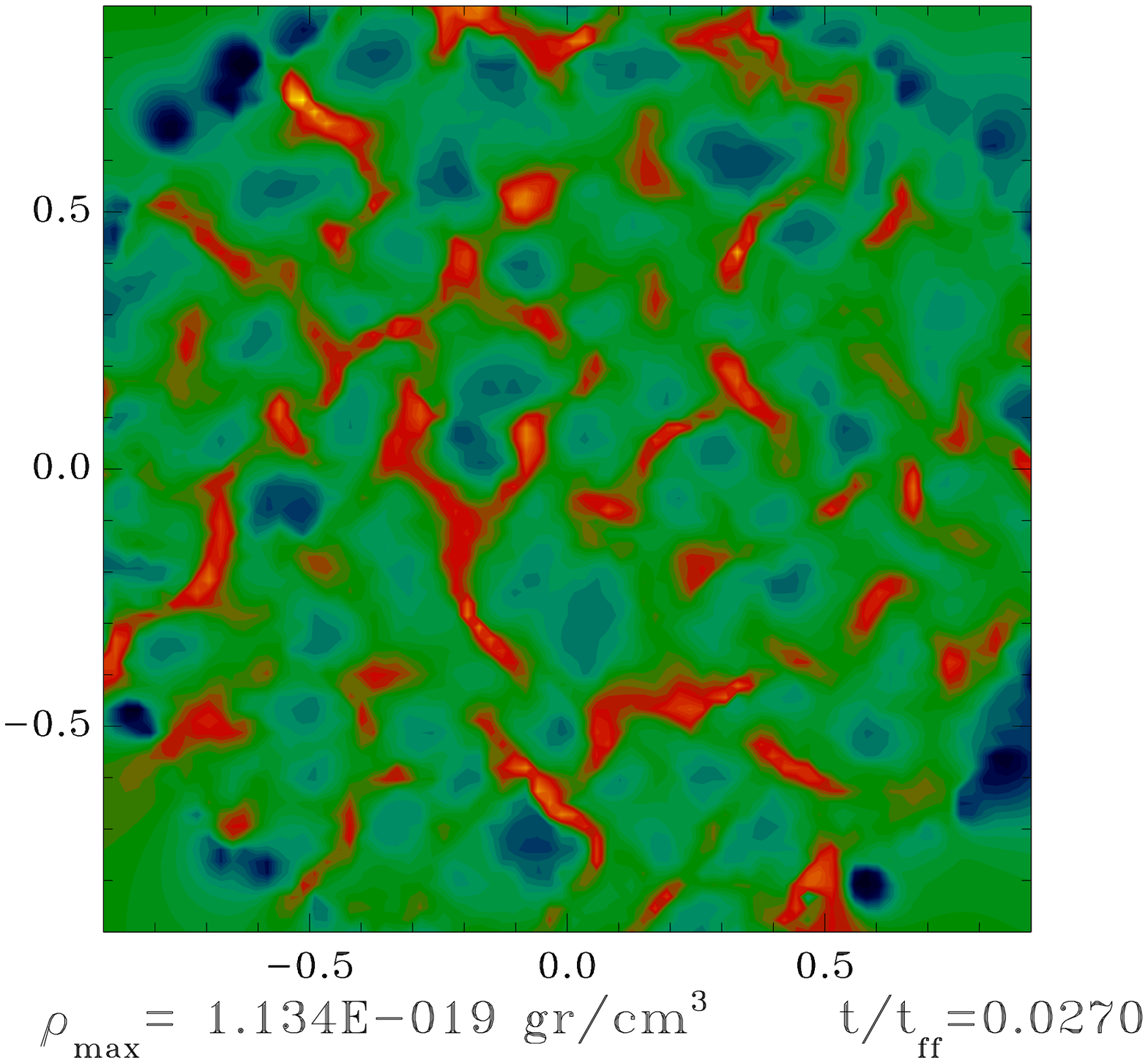} & \includegraphics[width=1.9 in,height=1.9 in]{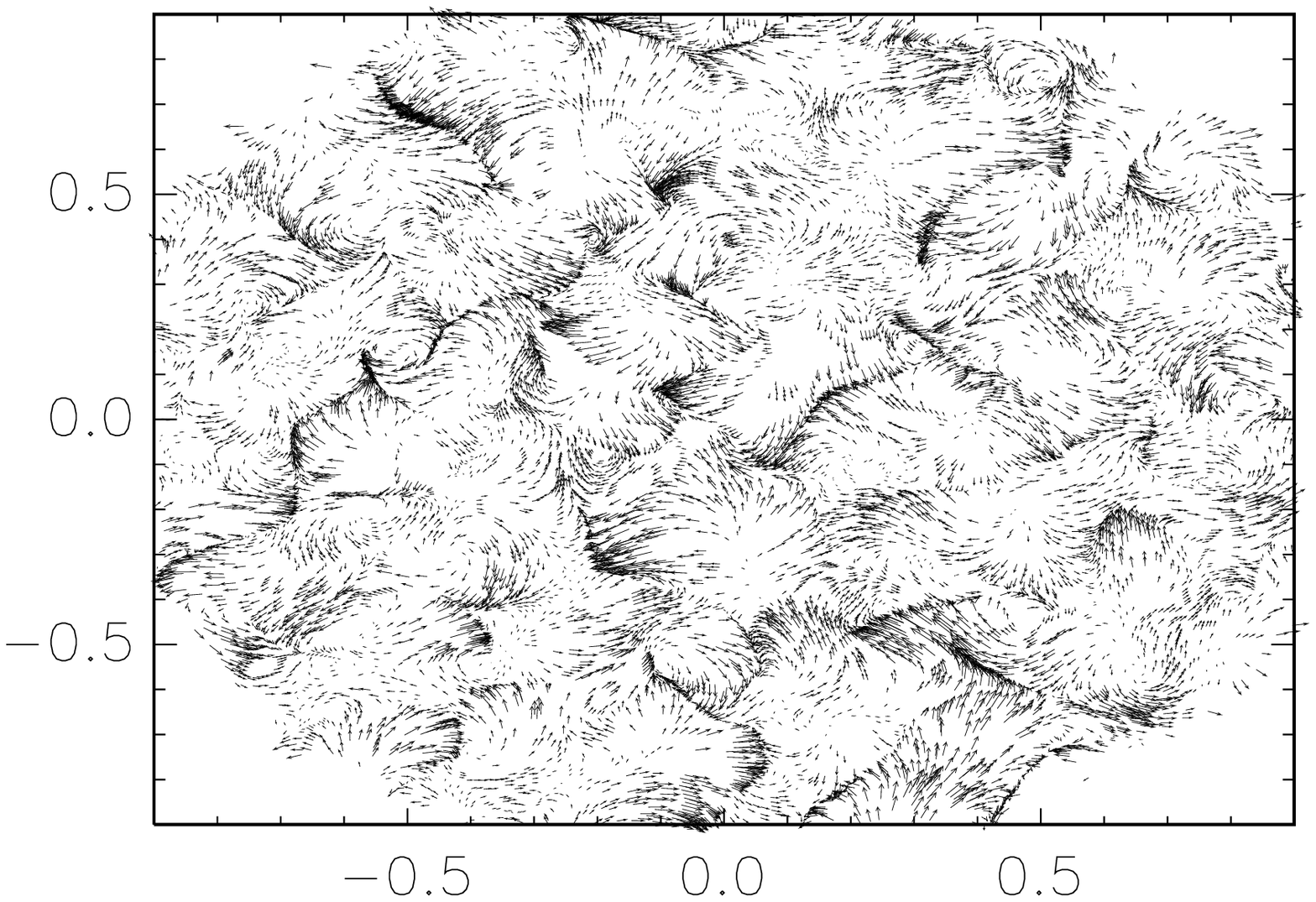}\\
\includegraphics[width=2.0in, height=2.0 in]{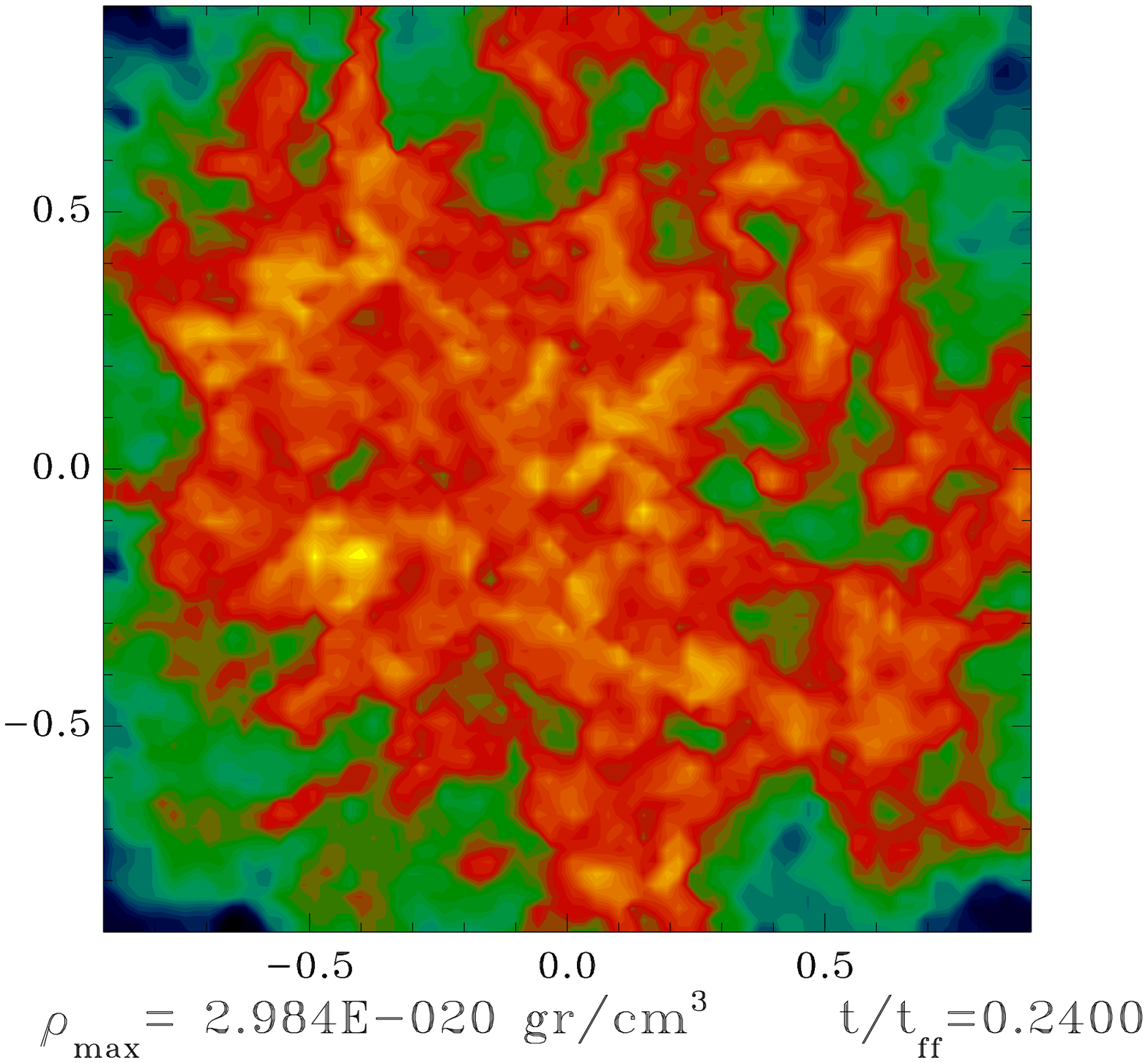}&\includegraphics[width=1.9 in,height=1.9 in]{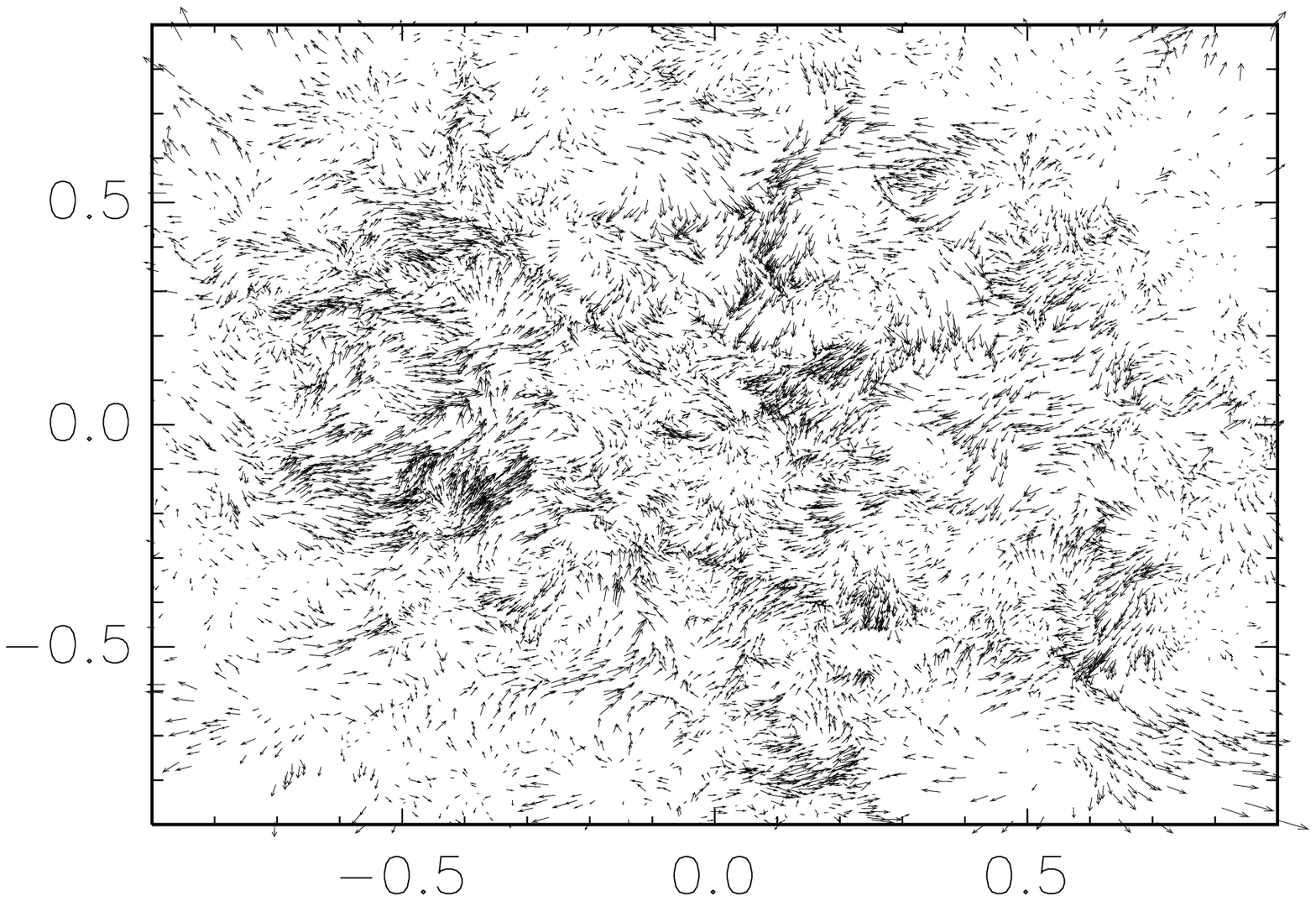}\\
\includegraphics[width=2.0in, height=2.0 in]{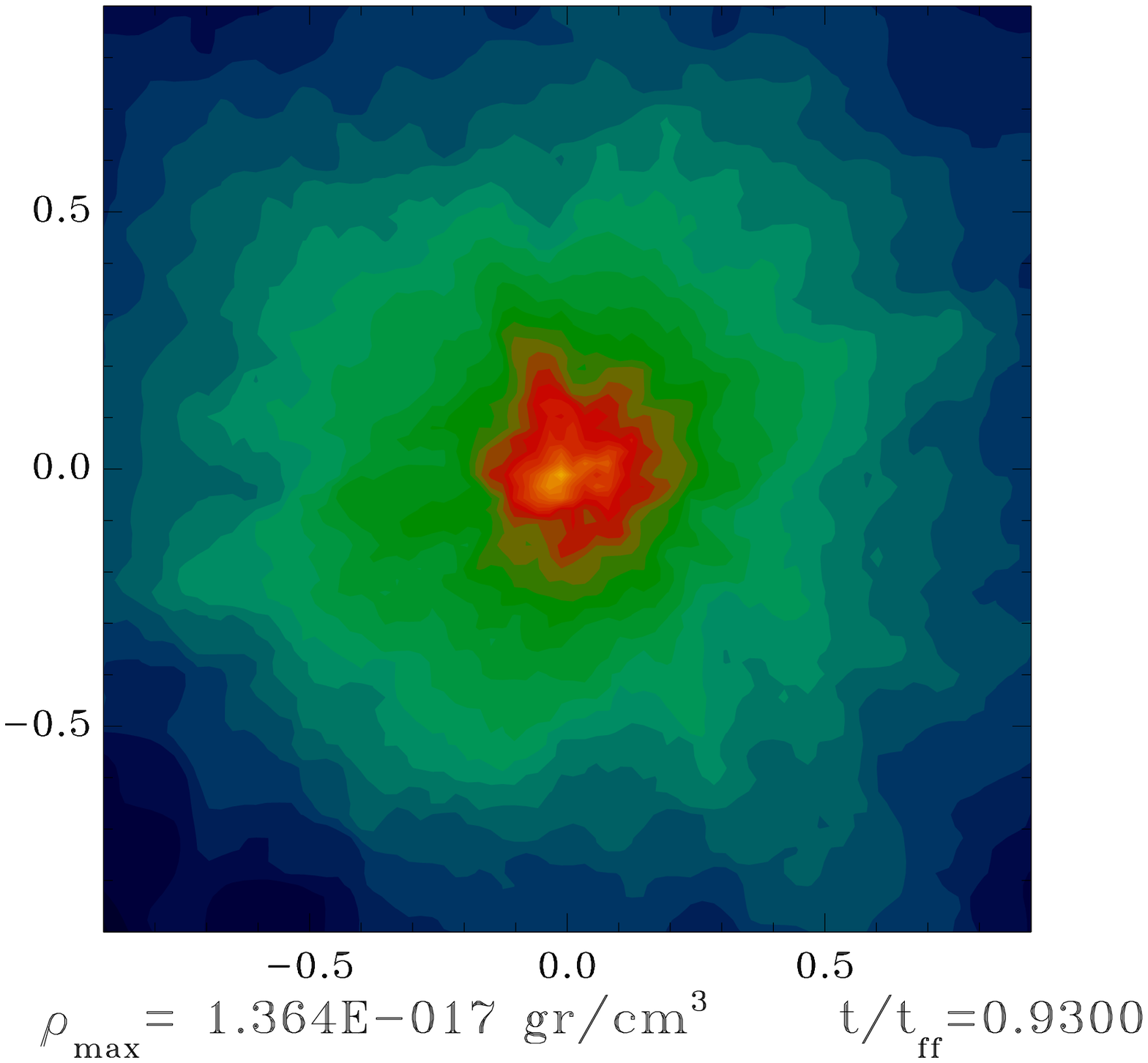}&\includegraphics[width=1.9 in,height=1.9 in]{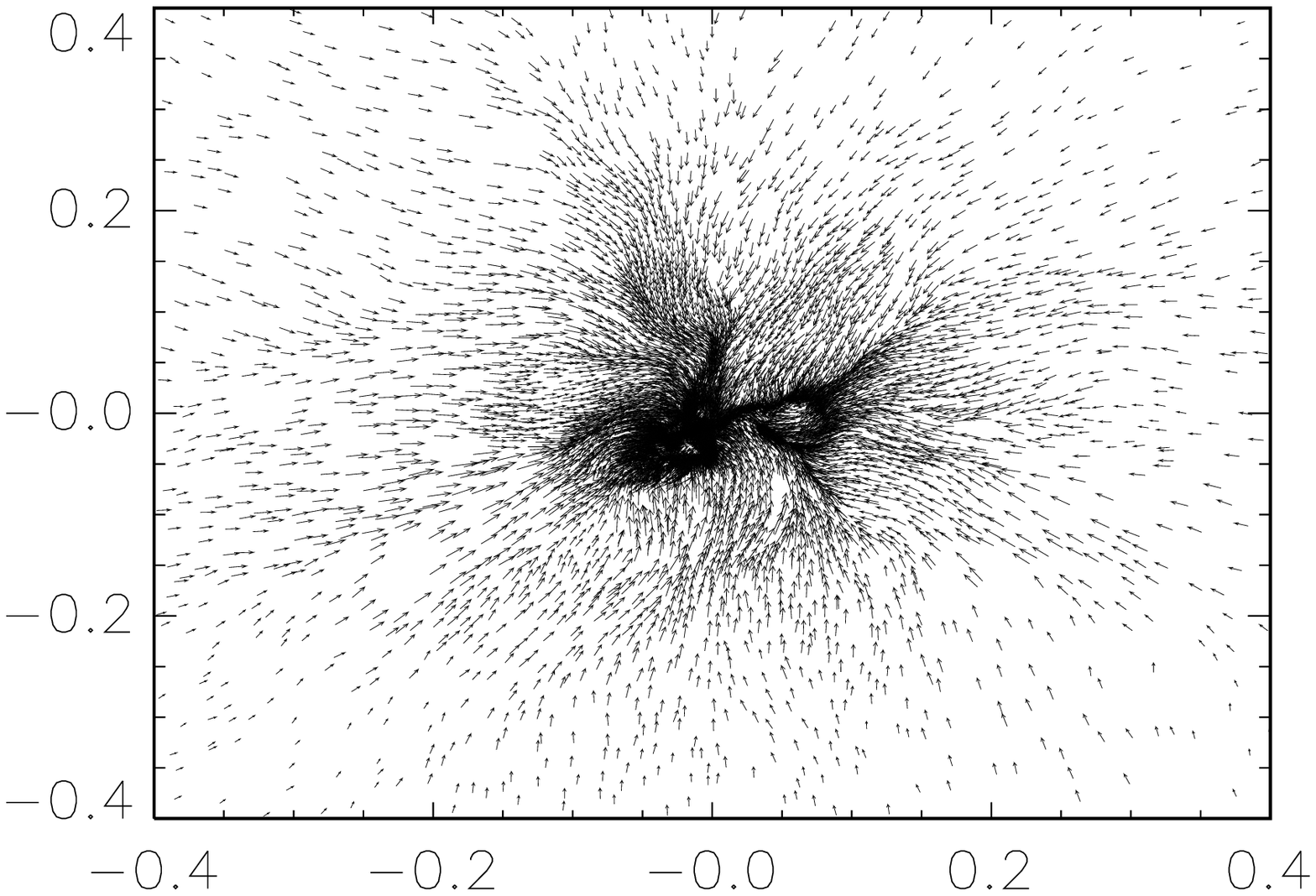}
\end{tabular}
\caption{ \label{ColColnpmR} Iso-density plots for the turbulent clump.}
\end{center}
\end{figure}

\subsection{The evolution of the turbulent clump.}
\label{ssec:evoturbclump}

One of the characteristics of turbulence is the appearance of a
filamentary and flocculent structure across the clump, a structure
which can already be seen in the first two panels of
Fig.\ref{ColColnpmR}. Because of the initial conditions chosen for
the clump, there is a clear tendency to a global collapse towards
its central region, as can be seen in the last panels of
Fig.\ref{ColColnpmR}.

We emphasize now a very important fact occurring at the outer
regions of the turbulent clump.  The turbulent clump is not in
hydrodynamic equilibrium and there is no external pressure acting
upon the clump, then the outermost particles have a non equilibrated
thermal pressure. Therefore, the outer clump particles expand
outwards. So, we have to keep in mind this expansion effect for the
problem at hand, as we shall quantify the mass of the clump swept
out by the winds.

\begin{figure}
\begin{center}
\begin{tabular}{ccc}
\hspace{-1 cm } \includegraphics[width=2.5 in]{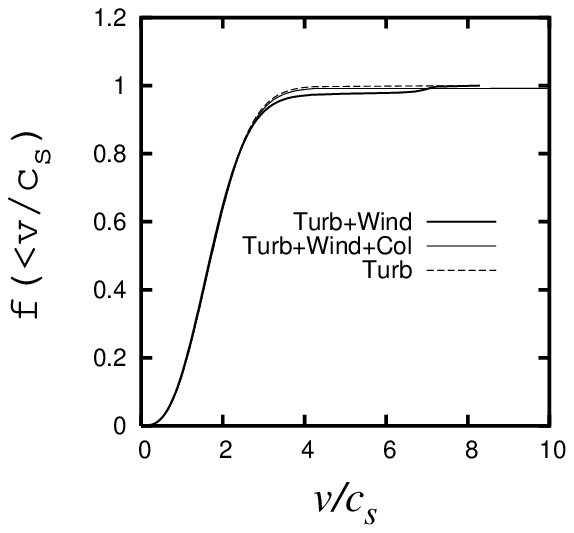}& \hspace{-1 cm } \includegraphics[width=2.5 in]{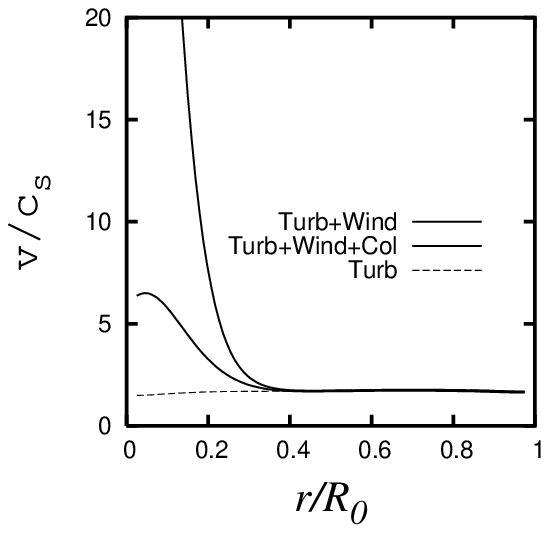} & \hspace{-1 cm } \includegraphics[width=2.5 in]{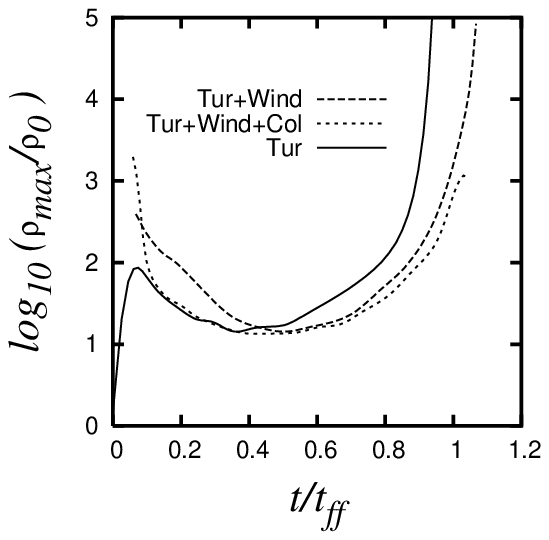}\\
\end{tabular}
\caption{ \label{velturviento} (left) the velocity distribution of the
particles; (middle) the velocity radial
profile and (right) the peak density time evolution.}
\end{center}
\end{figure}

\subsection{The effects of winds in the evolution of the turbulent clump.}
\label{ssec:effewindsturbclump}

As one can see in the two first panels of
Fig.~\ref{velturviento}, a very small fraction of gas particles can attain velocities
much higher than those velocities provided from the turbulence
alone, which are around $2$ Mach. Eighty per cent of the simulation particles have velocities
less than $v/cs<4$.

The winds are suddenly activated at the time $t/t_{ff}=0.05$, when the
clump has already acquired a fully flocculent aspect, which is a consequence of the
huge number of gas collisions produced randomly across the clump. This
time also marks the occurrence of the
highest peak in the clump's density curve, shown in the third panel of Fig.\ref{velturviento}

Besides, we notice by looking at the third panel
Fig.~\ref{velturviento}, that the global collapse of the clump does
not change even when the winds are introduced, as the peak density
curve of each run goes to higher values. However, the wind of the first
kind makes the collapse take place slower
as its density peak curve shows less steepness in the
middle stages of its evolution.

\begin{figure}
\begin{center}
\begin{tabular}{cc}
\includegraphics[width=2.2 in, height=2.2 in]{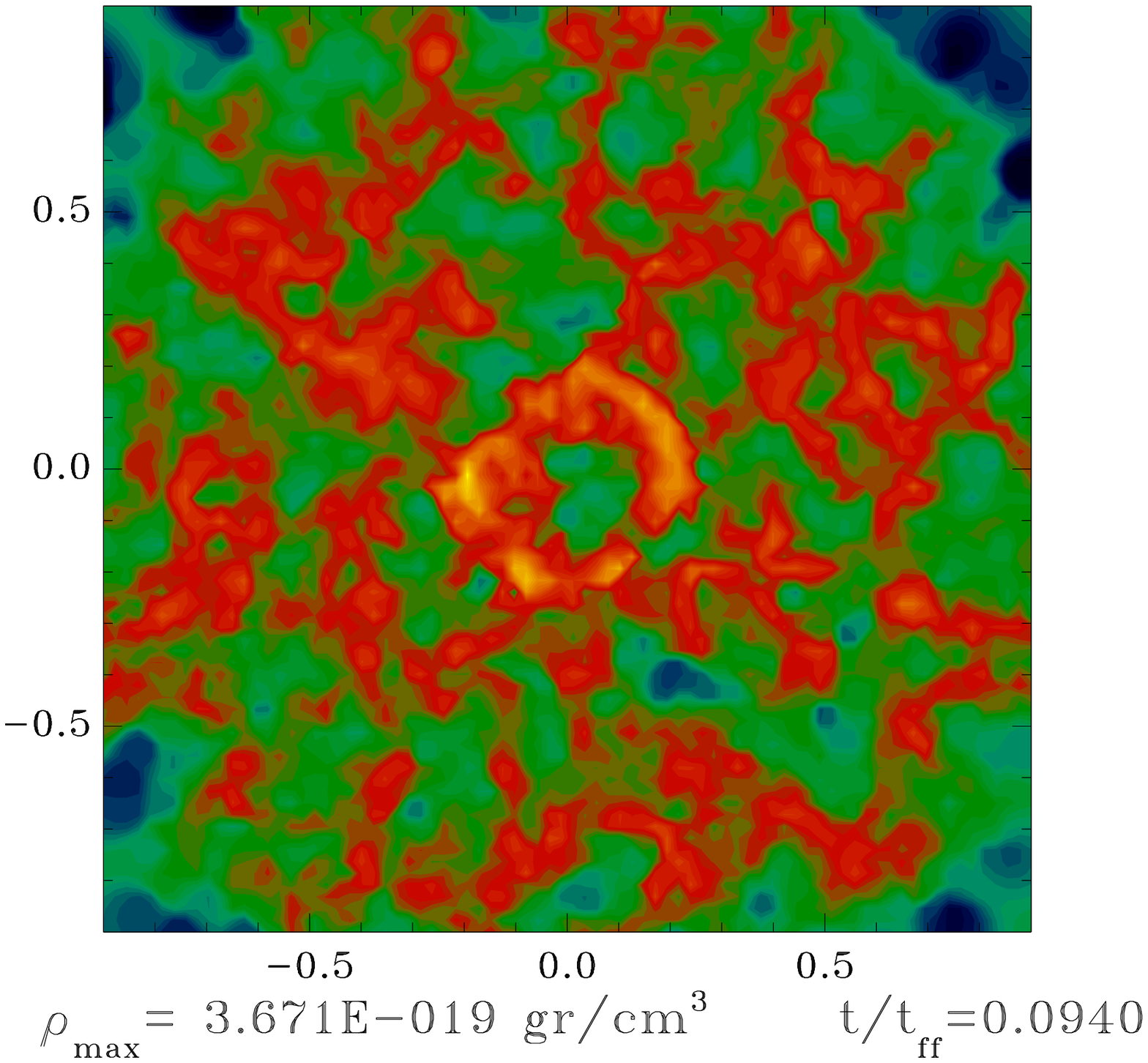}&\includegraphics[width=2.15 in,height=2.15 in]{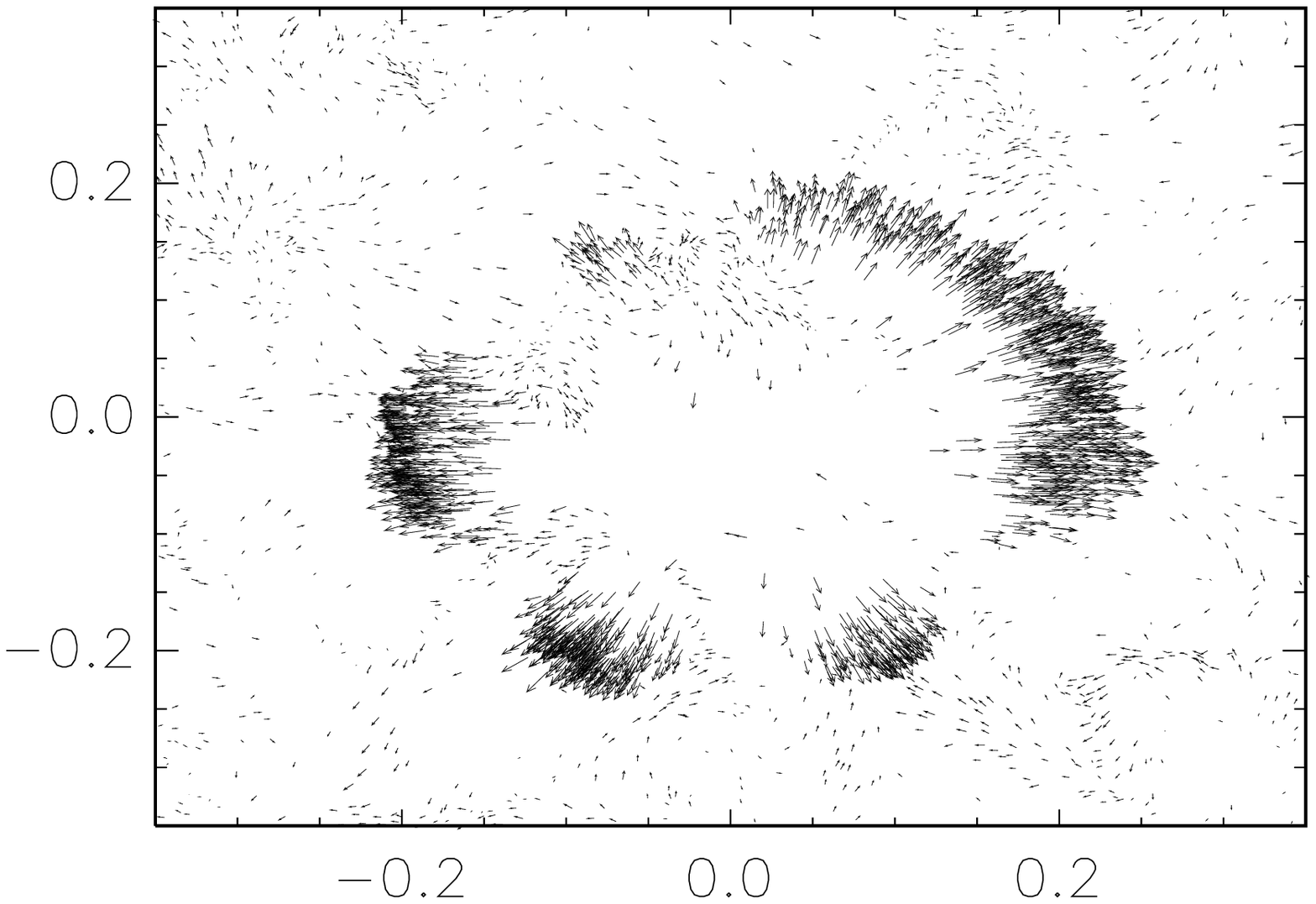}\\
\includegraphics[width=2.2 in, height=2.2 in]{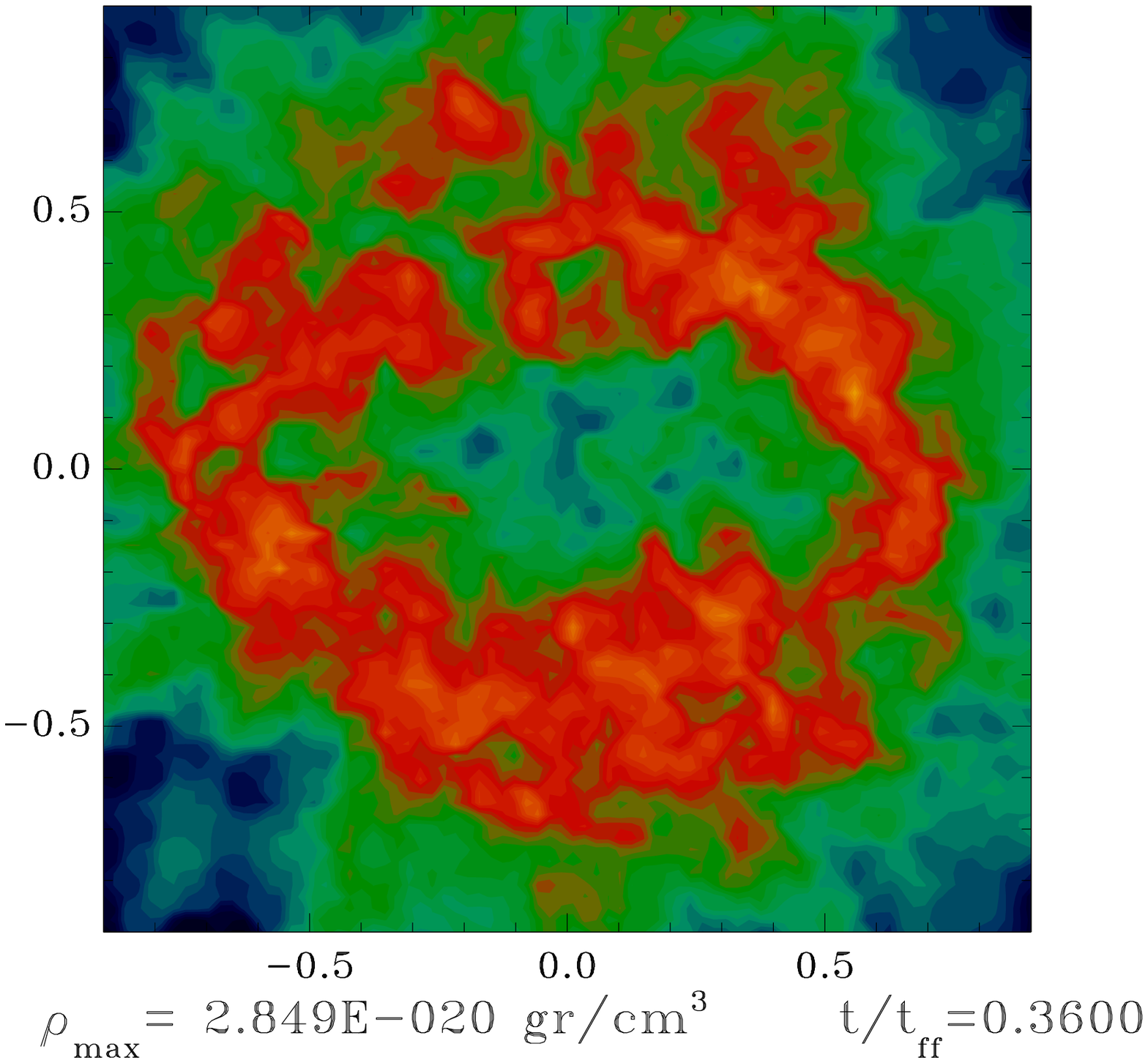}&\includegraphics[width=2.15 in,height=2.15 in]{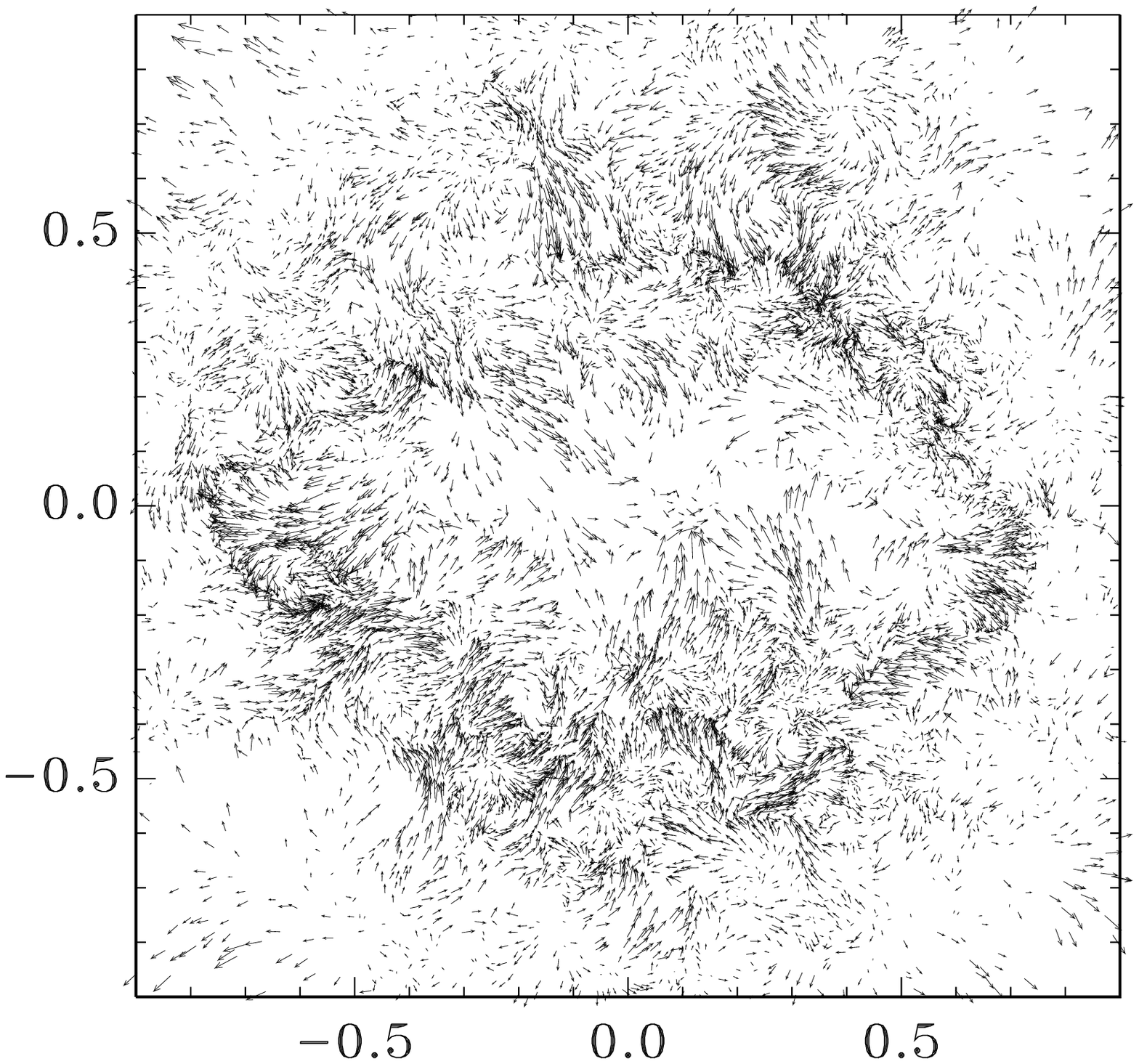}\\
\includegraphics[width=2.2 in, height=2.2 in]{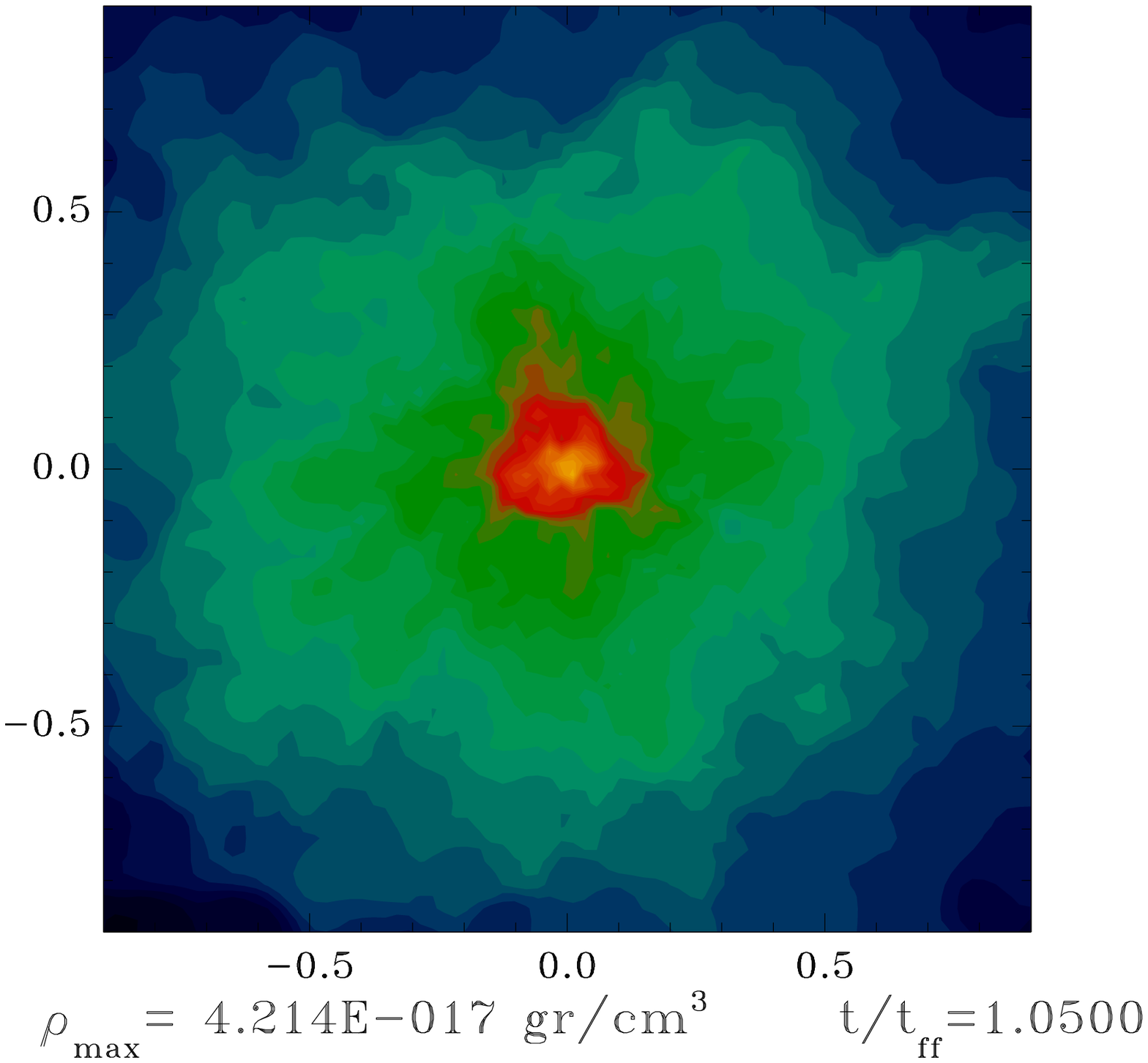}&\includegraphics[width=2.15 in,height=2.15 in]{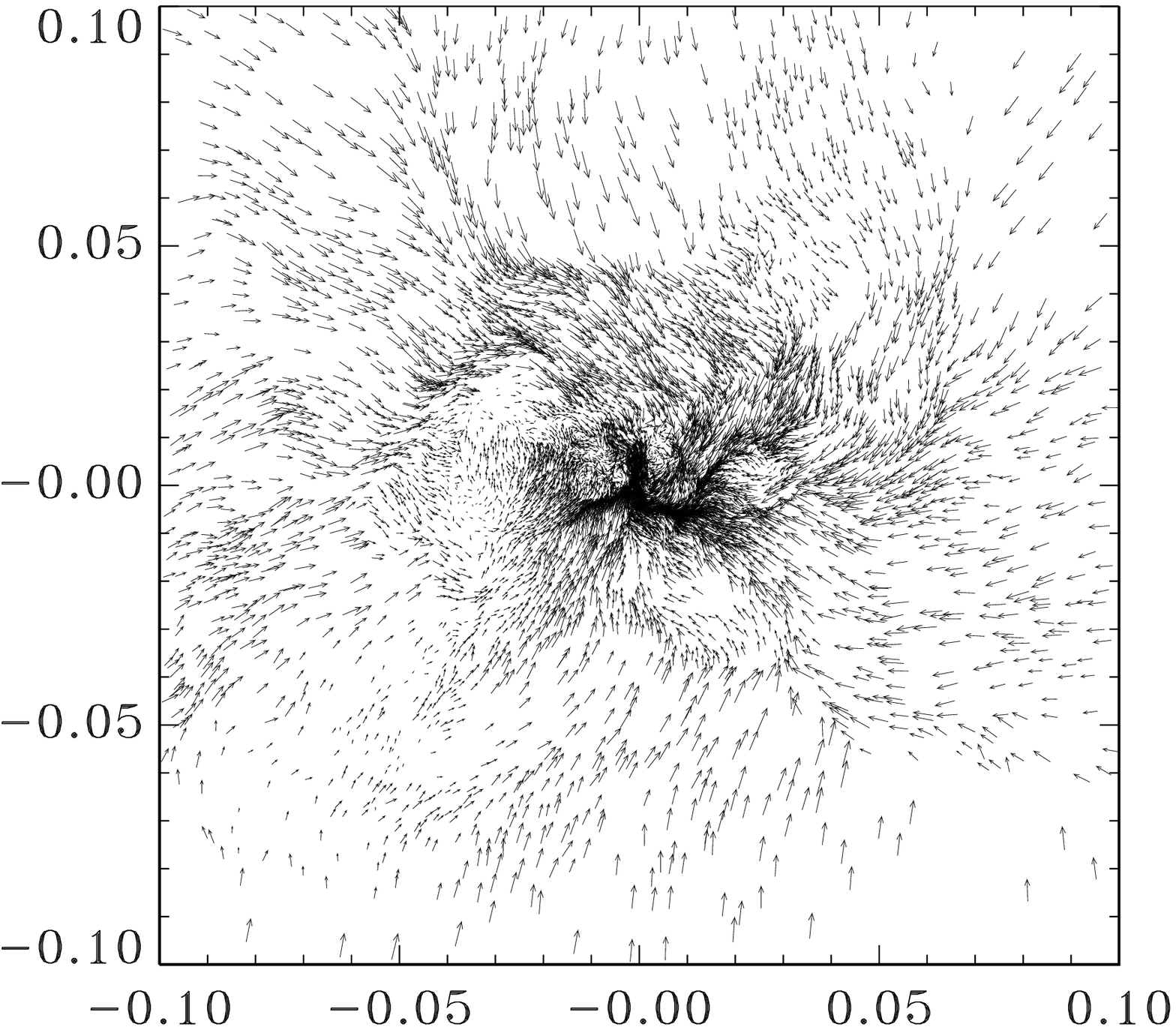}\\
\end{tabular}
\caption{ \label{NSTVRR6pp} Iso-density and velocity plots for the model "Tur+Wind".}
\end{center}
\end{figure}

The iso-density plots for the simulation $Tur+Wind$ are shown in
Fig.~\ref{NSTVRR6pp}.  By the time $t/t_{ff}=0.36$, we see a void
created in the central region of the clump because
both the wind particles and those particles which are swept out move
towards the outer parts of the clump. However, gravity and viscosity act together in such a
way that the particles quickly fill the void, as can be seen in
the last panel of Fig.~\ref{NSTVRR6pp}.

\begin{figure}
\begin{center}
\begin{tabular}{cc}
\includegraphics[width=2.2 in, height=2.2 in]{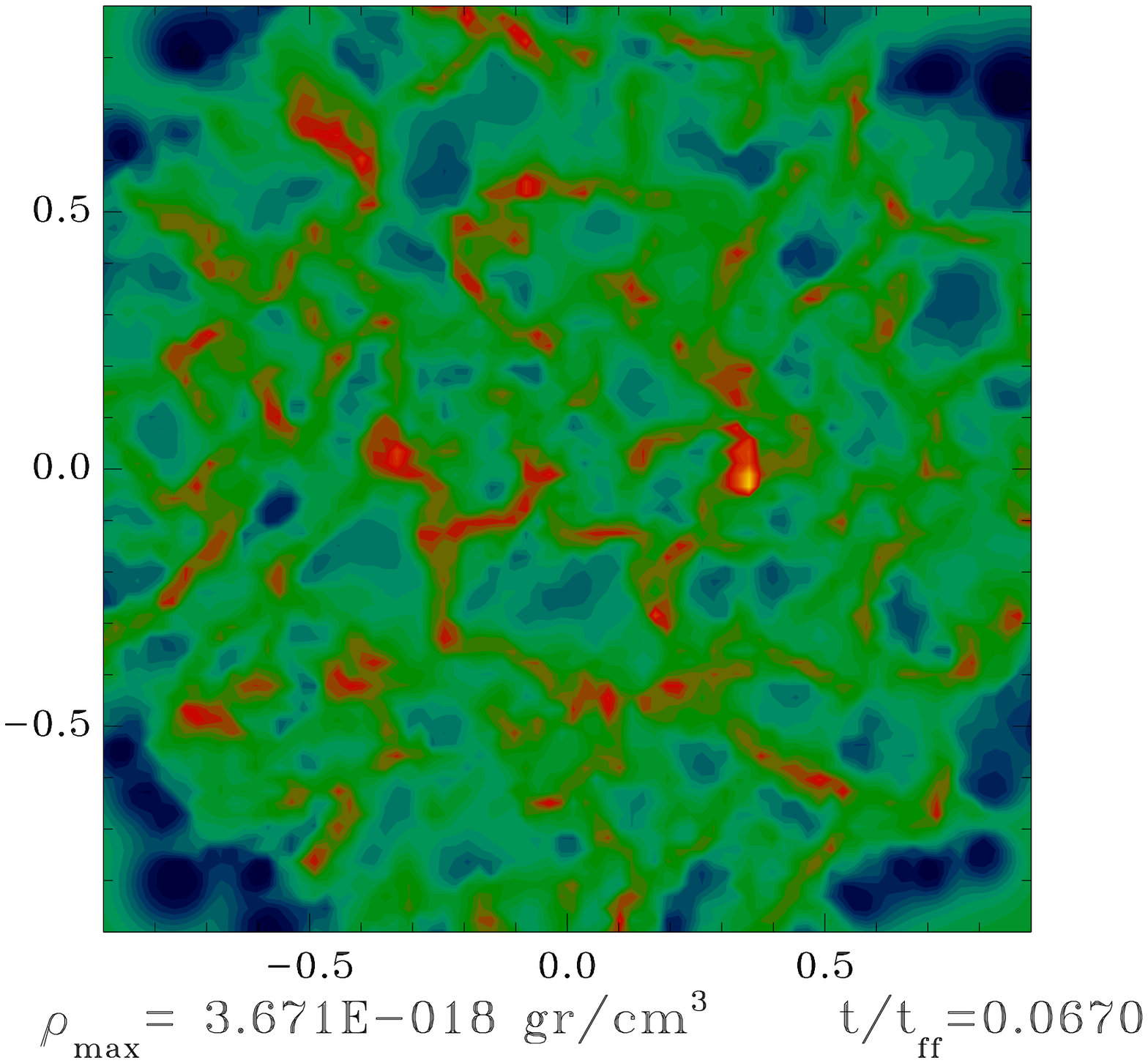}&\includegraphics[width=2.15 in,height=2.15 in]{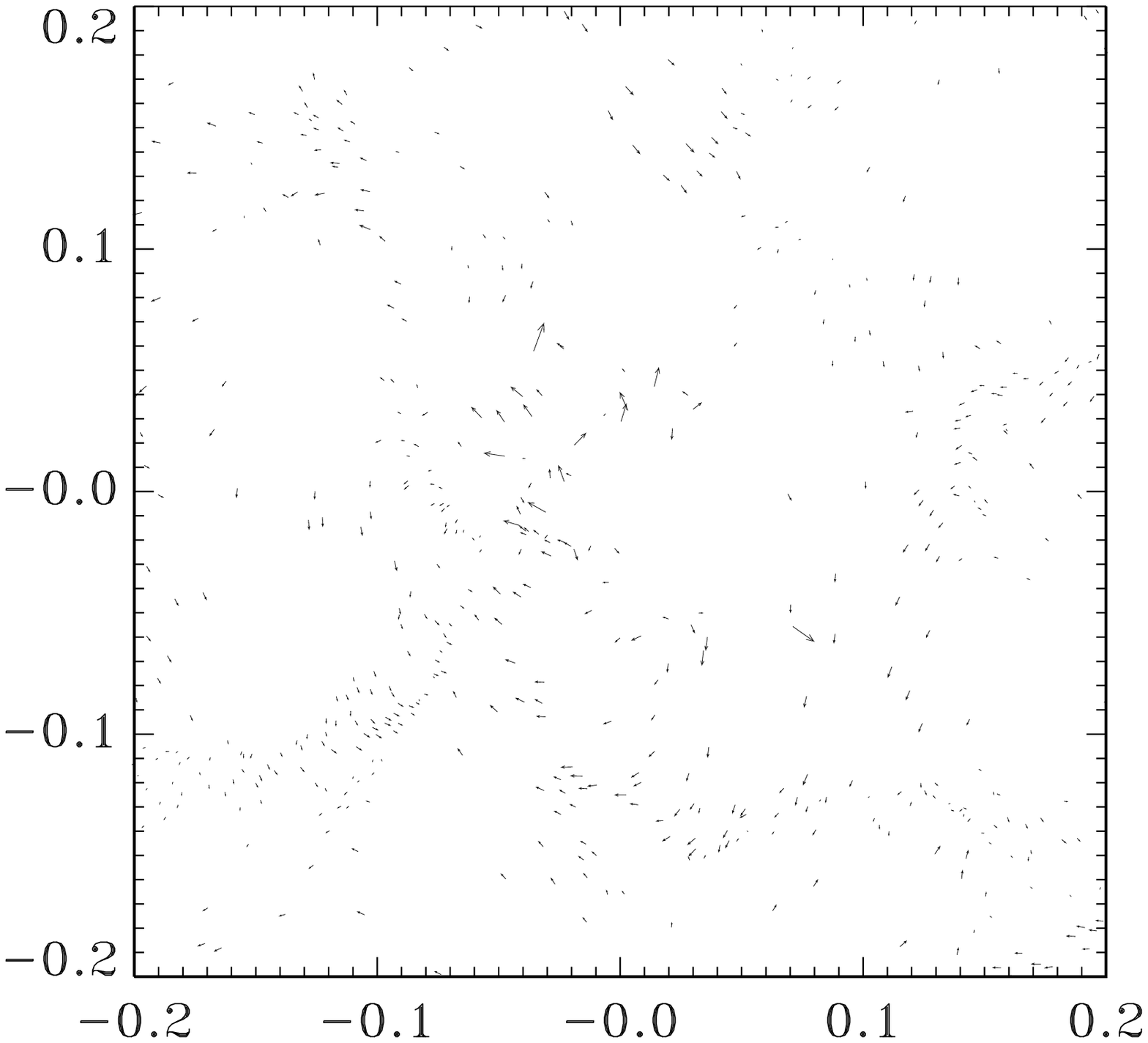}\\
\includegraphics[width=2.2 in, height=2.2 in]{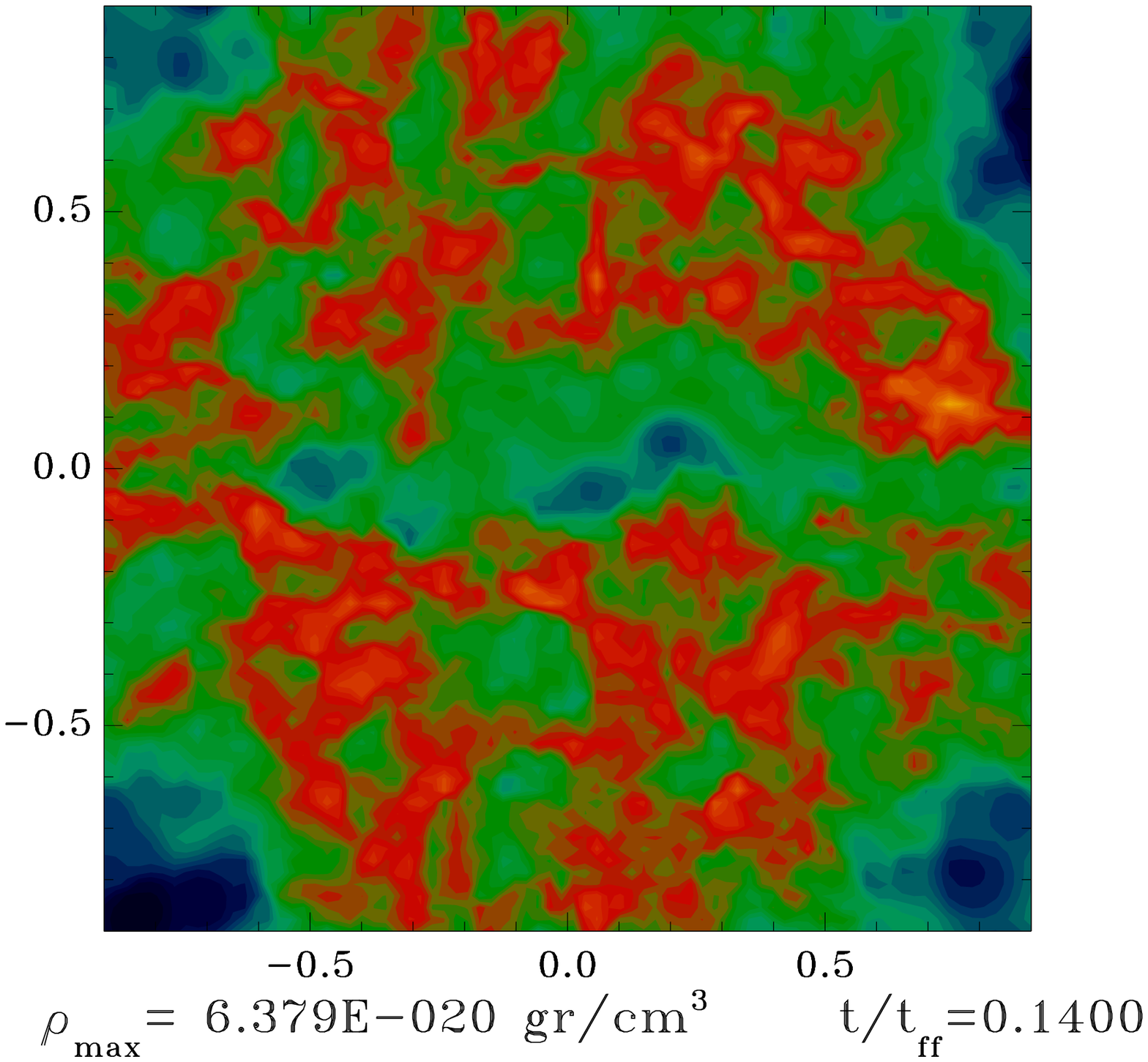}&\includegraphics[width=2.15 in,height=2.15 in]{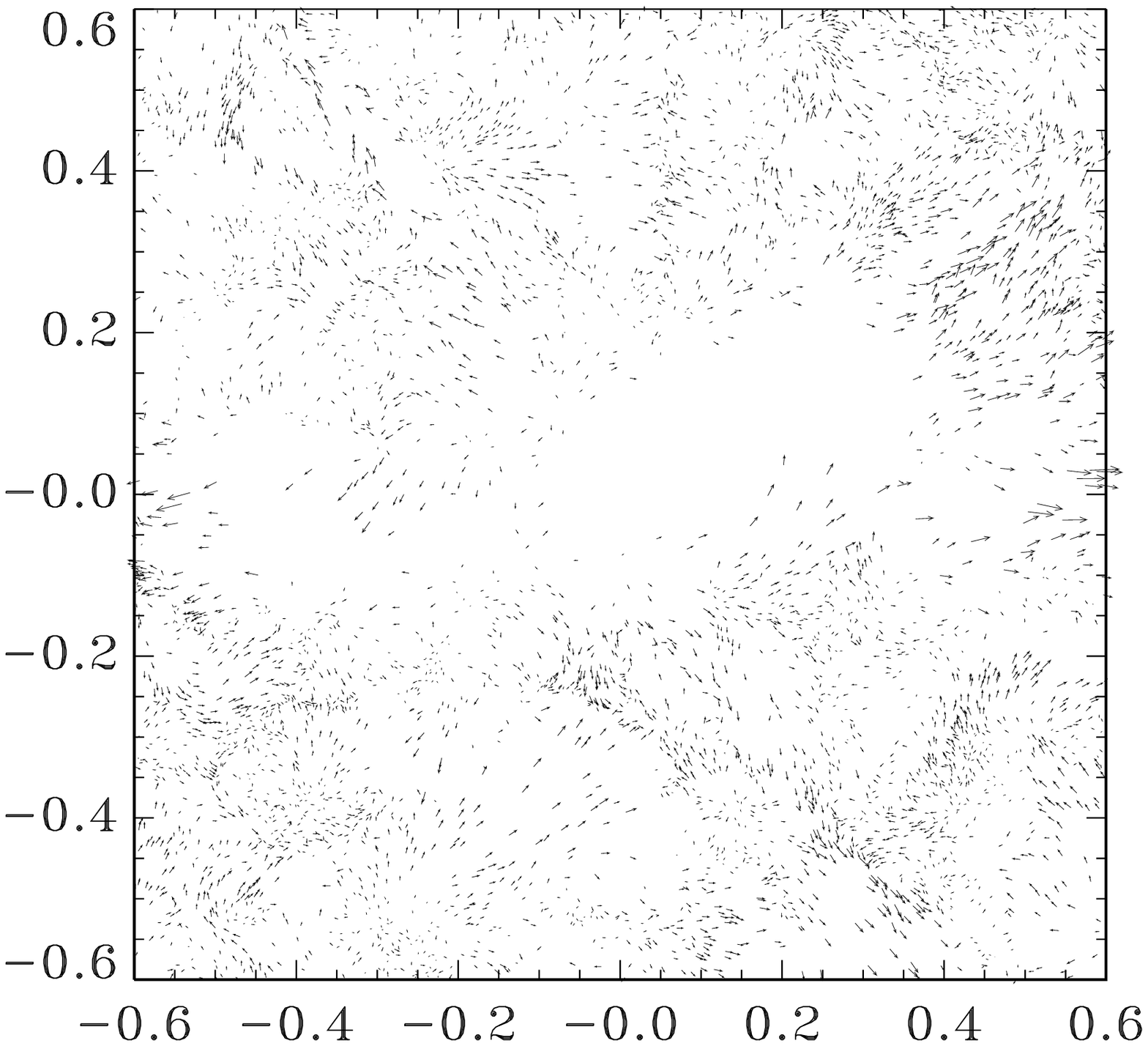}\\
\includegraphics[width=2.2 in, height=2.2 in]{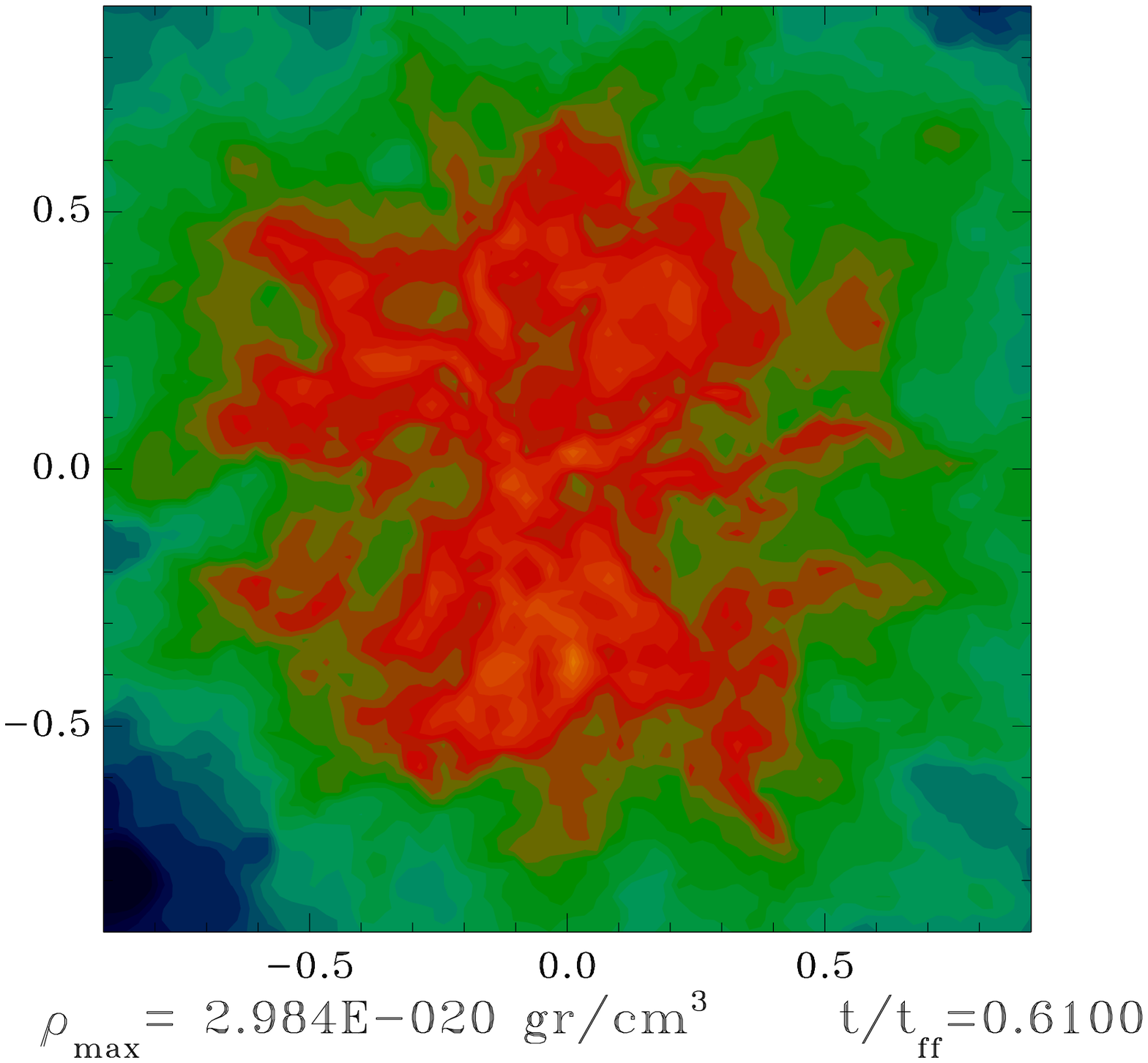}&\includegraphics[width=2.15 in,height=2.15 in]{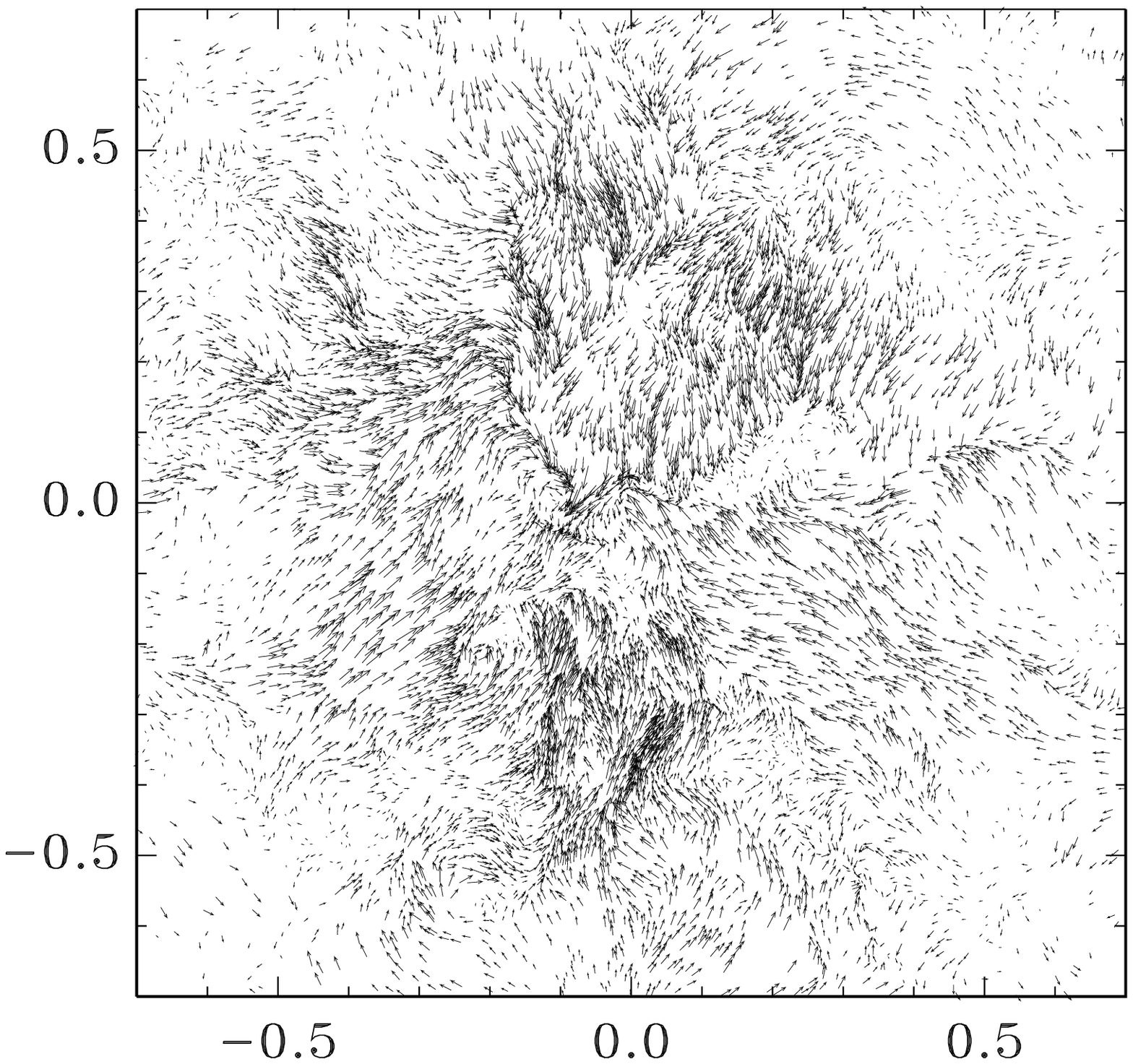}\\
\includegraphics[width=2.2 in, height=2.2 in]{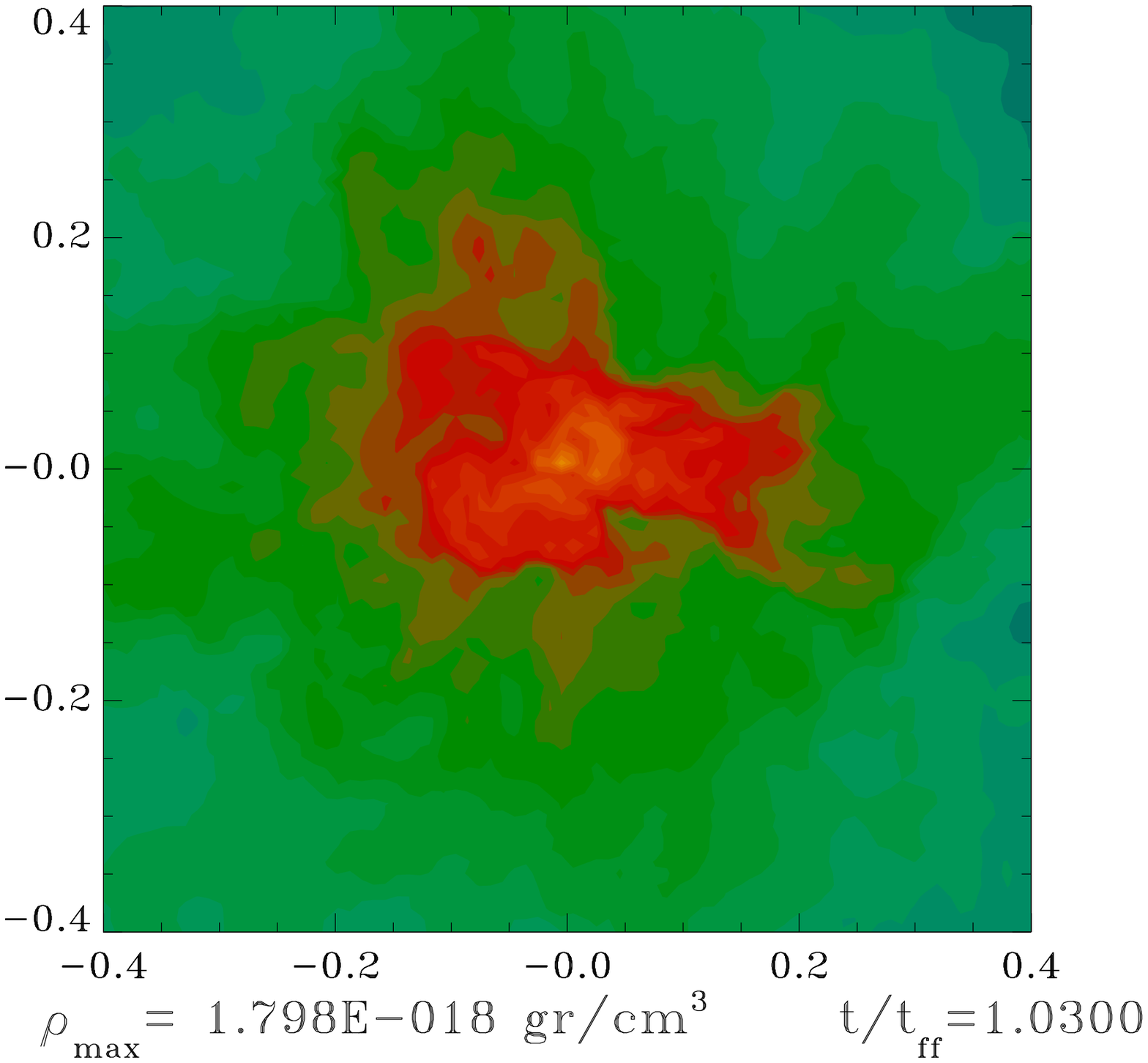}&\includegraphics[width=2.15 in,height=2.15 in]{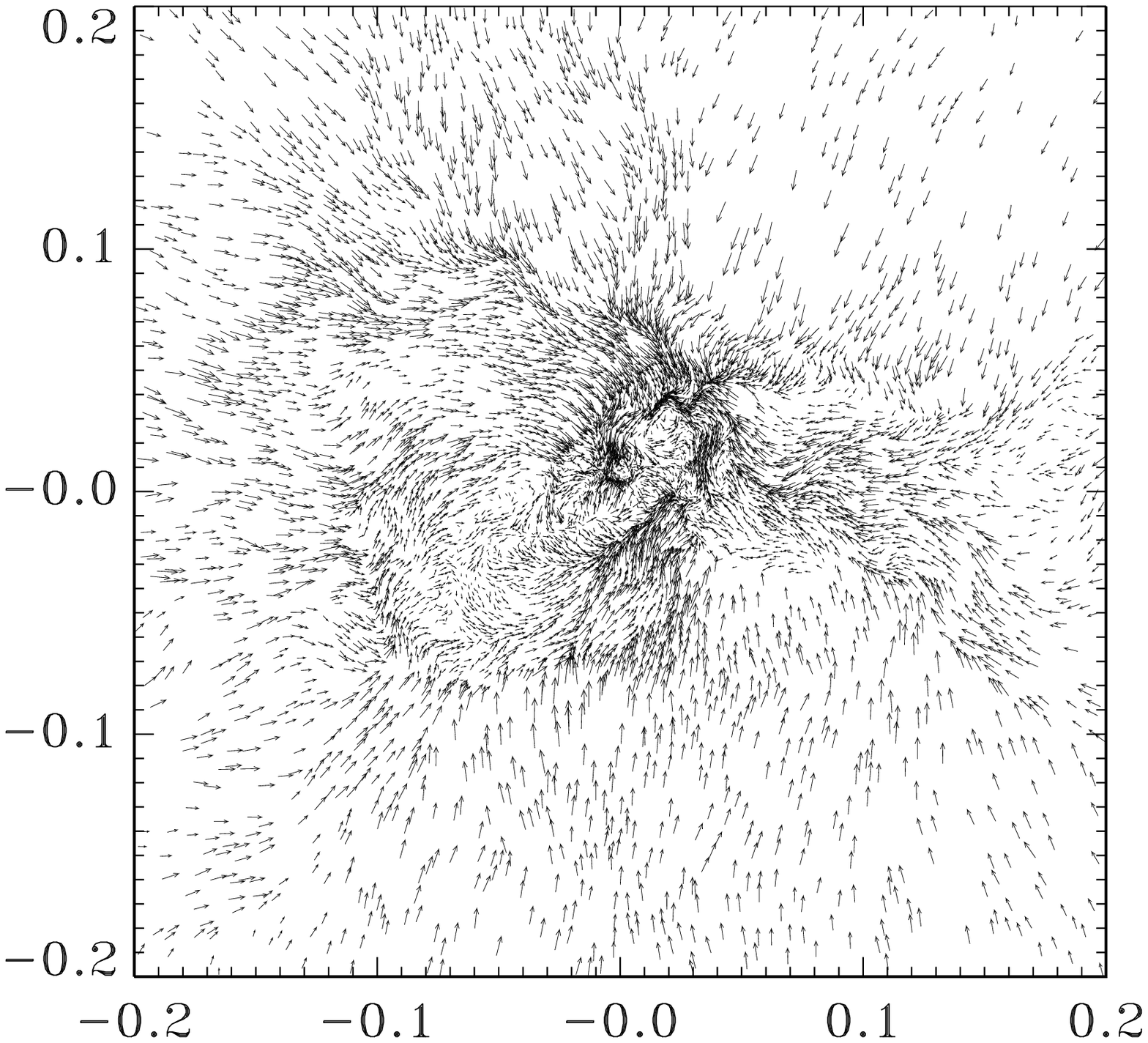}
\end{tabular}
\caption{\label{NSTVRRJ7pp} Iso-density and velocity plots for the model "Tur+Wind+Col".}
\end{center}
\end{figure}

When we consider the model  $Tur+Wind+Col$, so that a collimated gas of particles is
ejected, we see that the effects on the clump are more significant, but
essentially the same phenomena as seen in the model $Tur+Wind$ take place, as can
be seen in Fig.~\ref{NSTVRRJ7pp}.

\begin{figure}
\begin{center}
\begin{tabular}{cc}
\includegraphics[width=2.0 in]{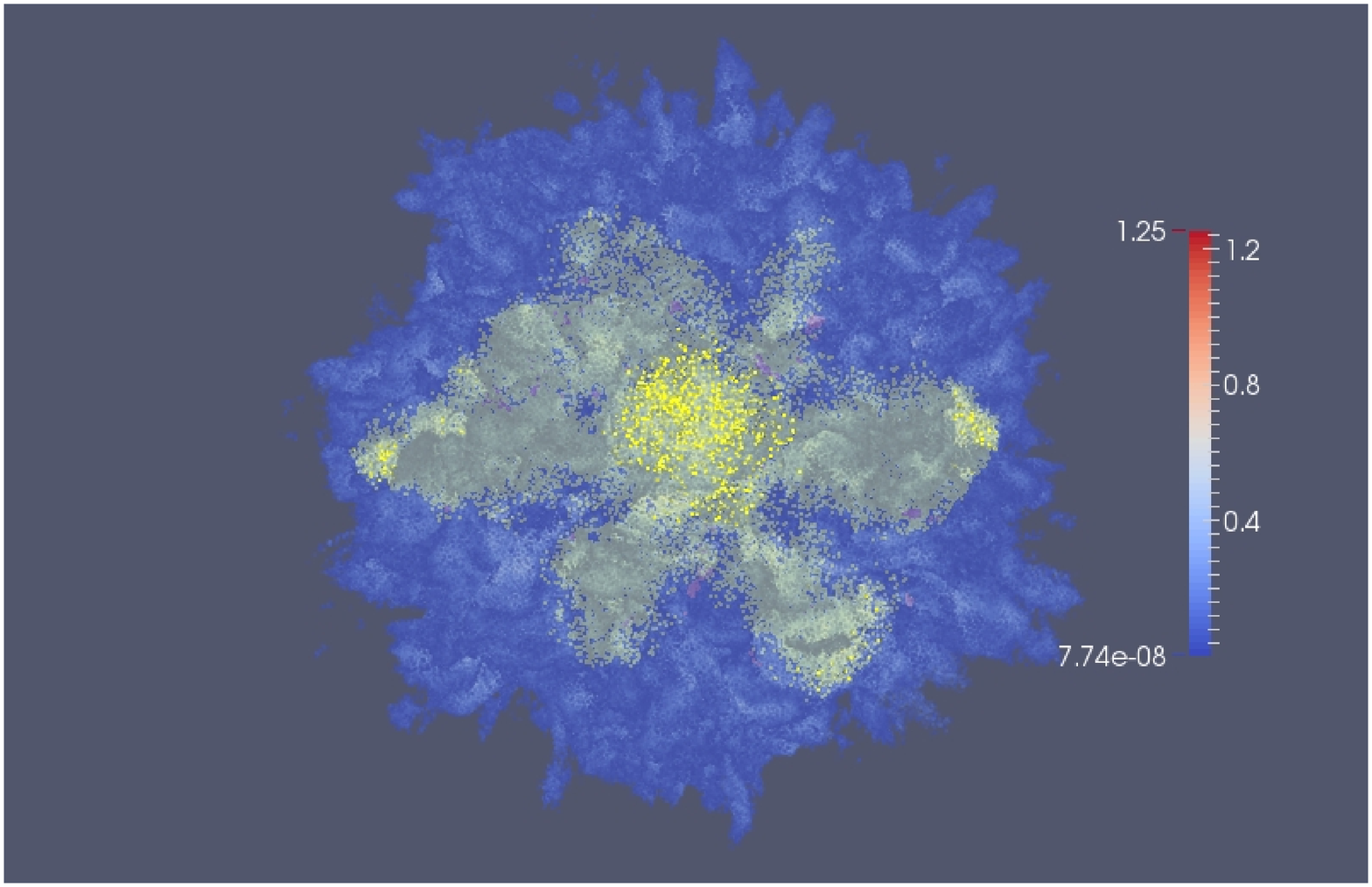} & \includegraphics[width=2.0 in]{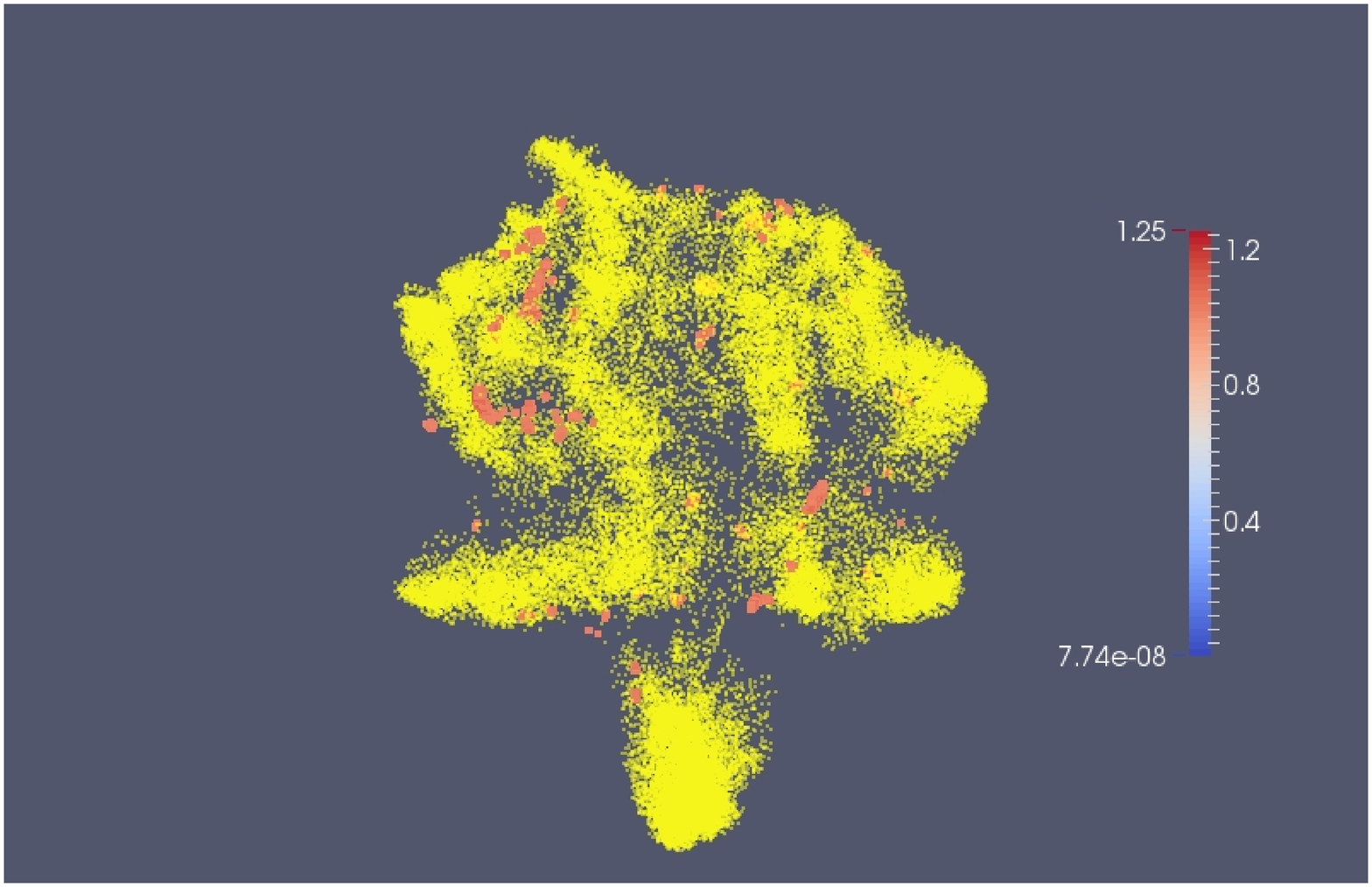} \\
\includegraphics[width=2.0 in]{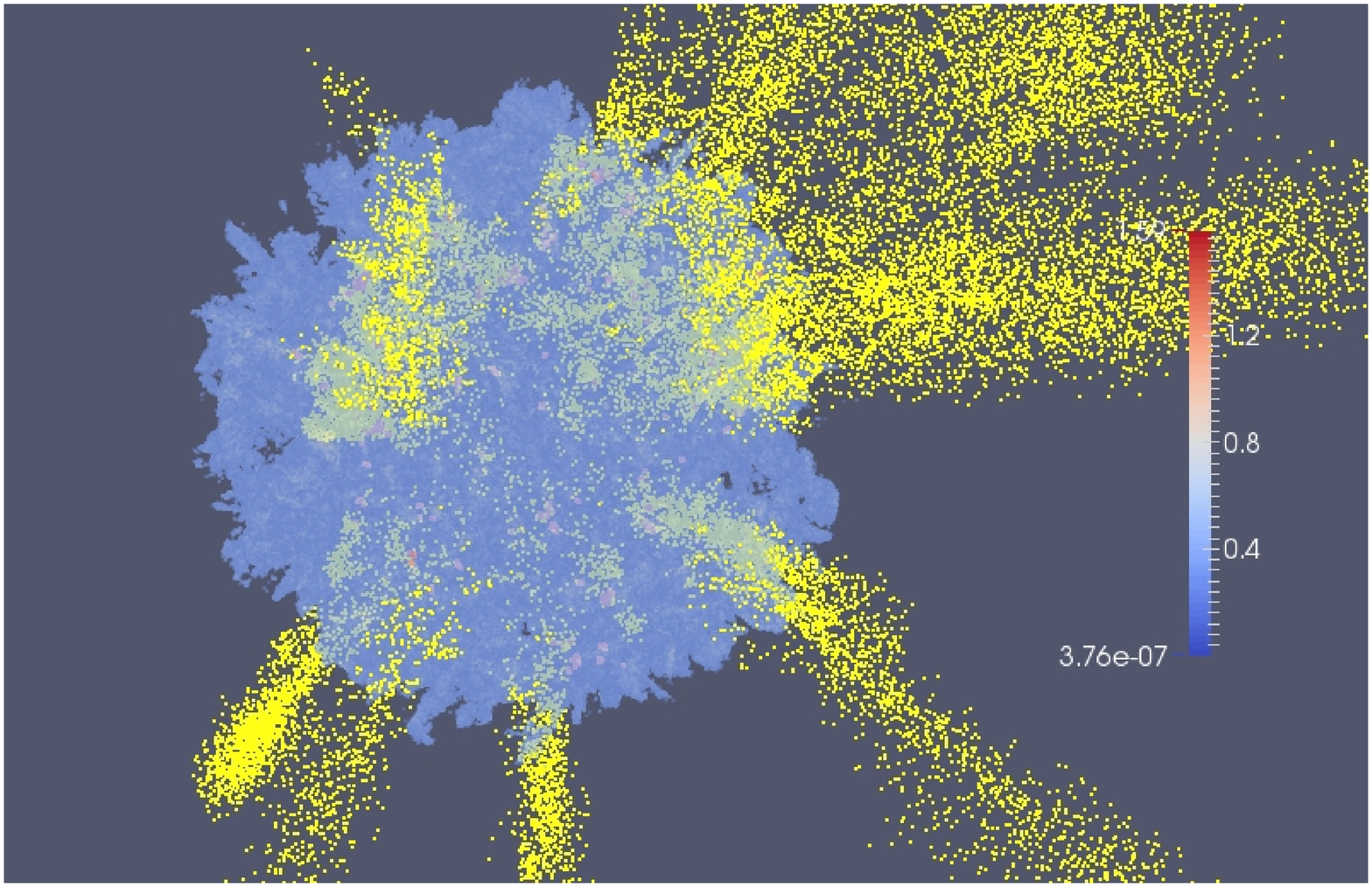} & \includegraphics[width=2.0 in]{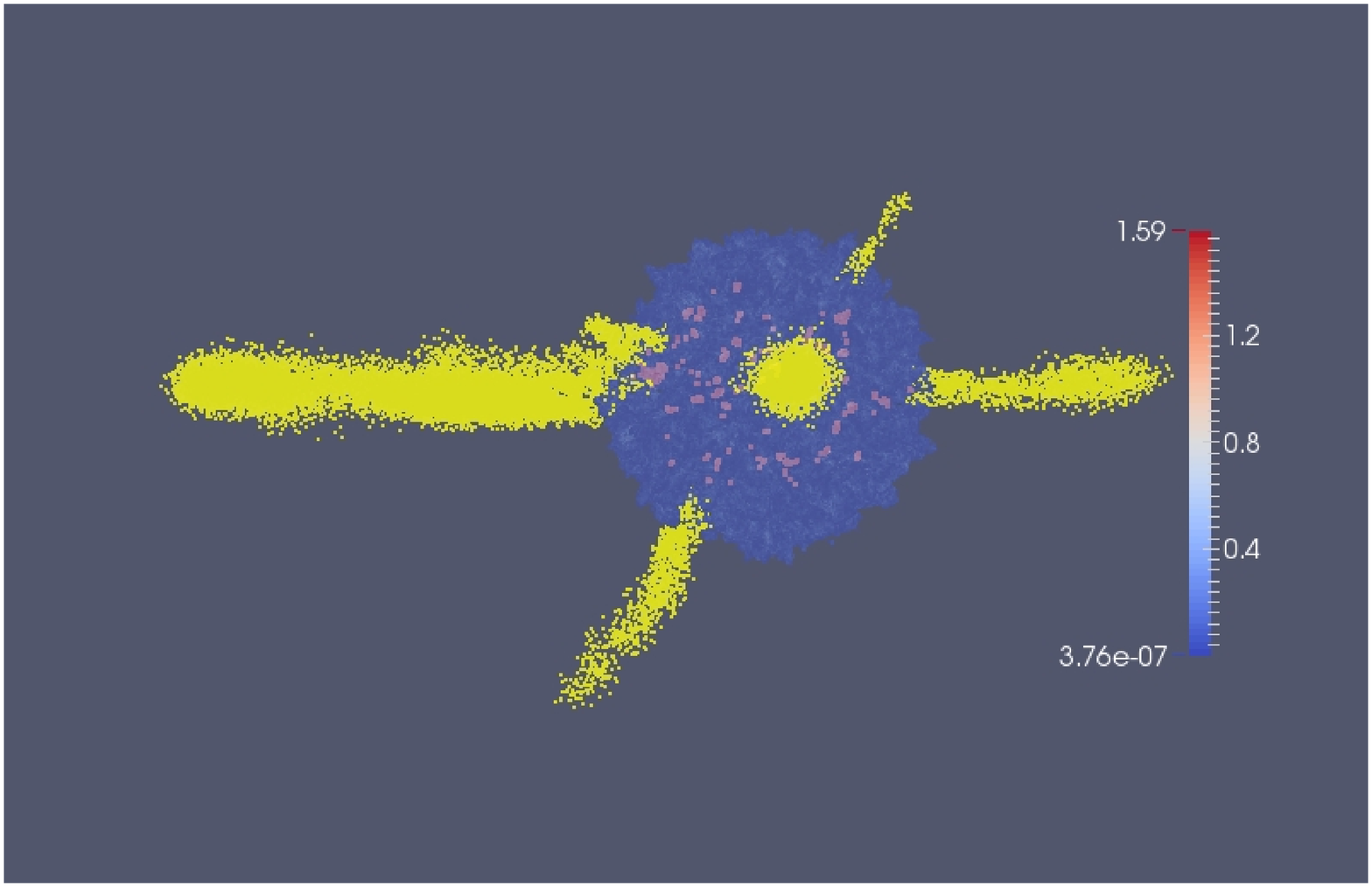} \\
\end{tabular}
\caption{ \label{vis3dGV} $3D$ plot of the turbulent clump (in blue) with winds (in
yellow) and formed accretion centers (in red) for models (top)
$Tur+Win$ and (bottom) $Tur+Win+Col$, corresponding to the plots
shown in the second line of Figs.\ref{NSTVRR6pp} and
\ref{NSTVRRJ7pp}, respectively.}
\end{center}
\end{figure}
\begin{figure}
\begin{center}
\begin{tabular}{cc}
\includegraphics[width=2.0 in]{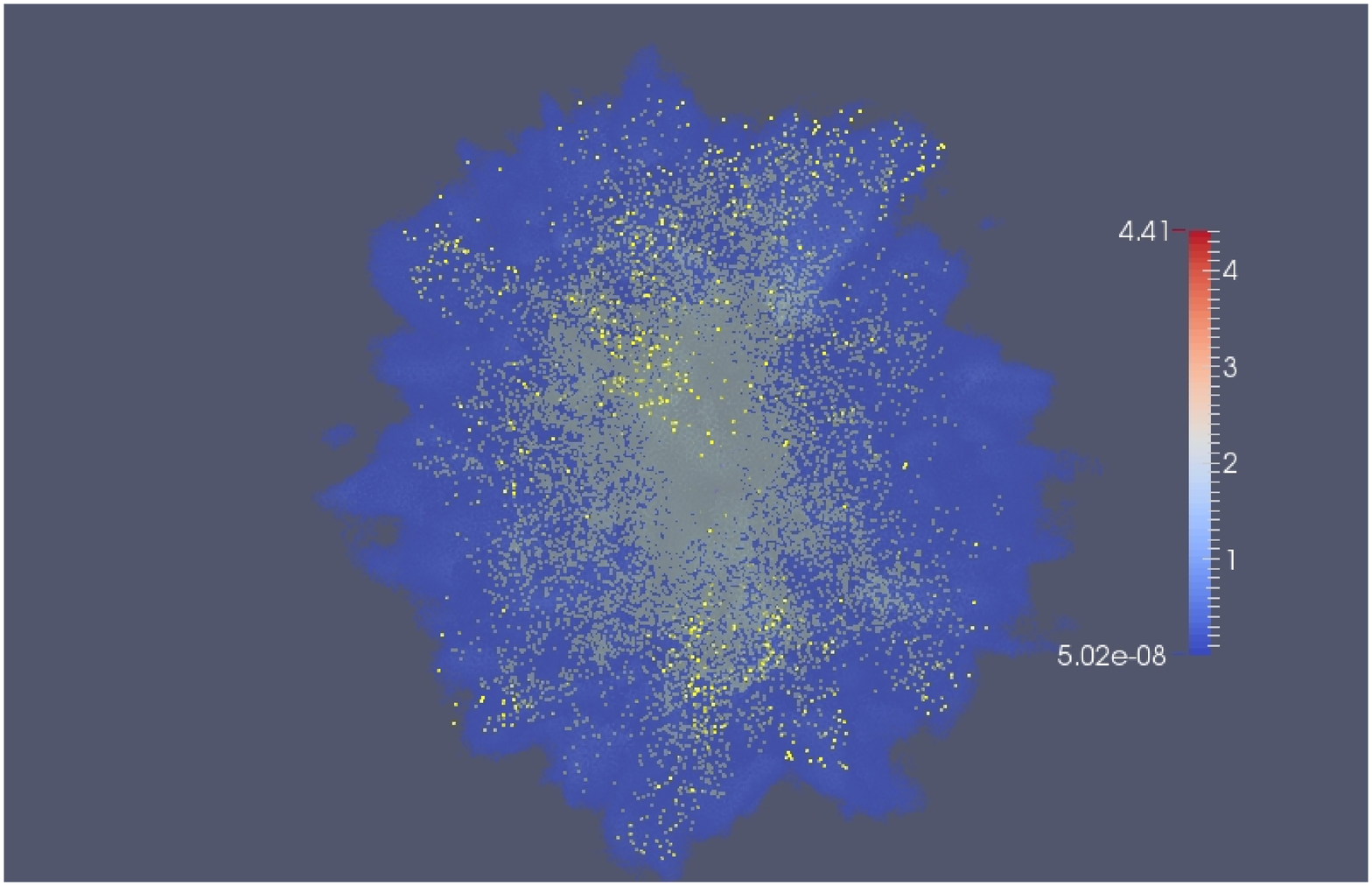} & \includegraphics[width=2.0 in]{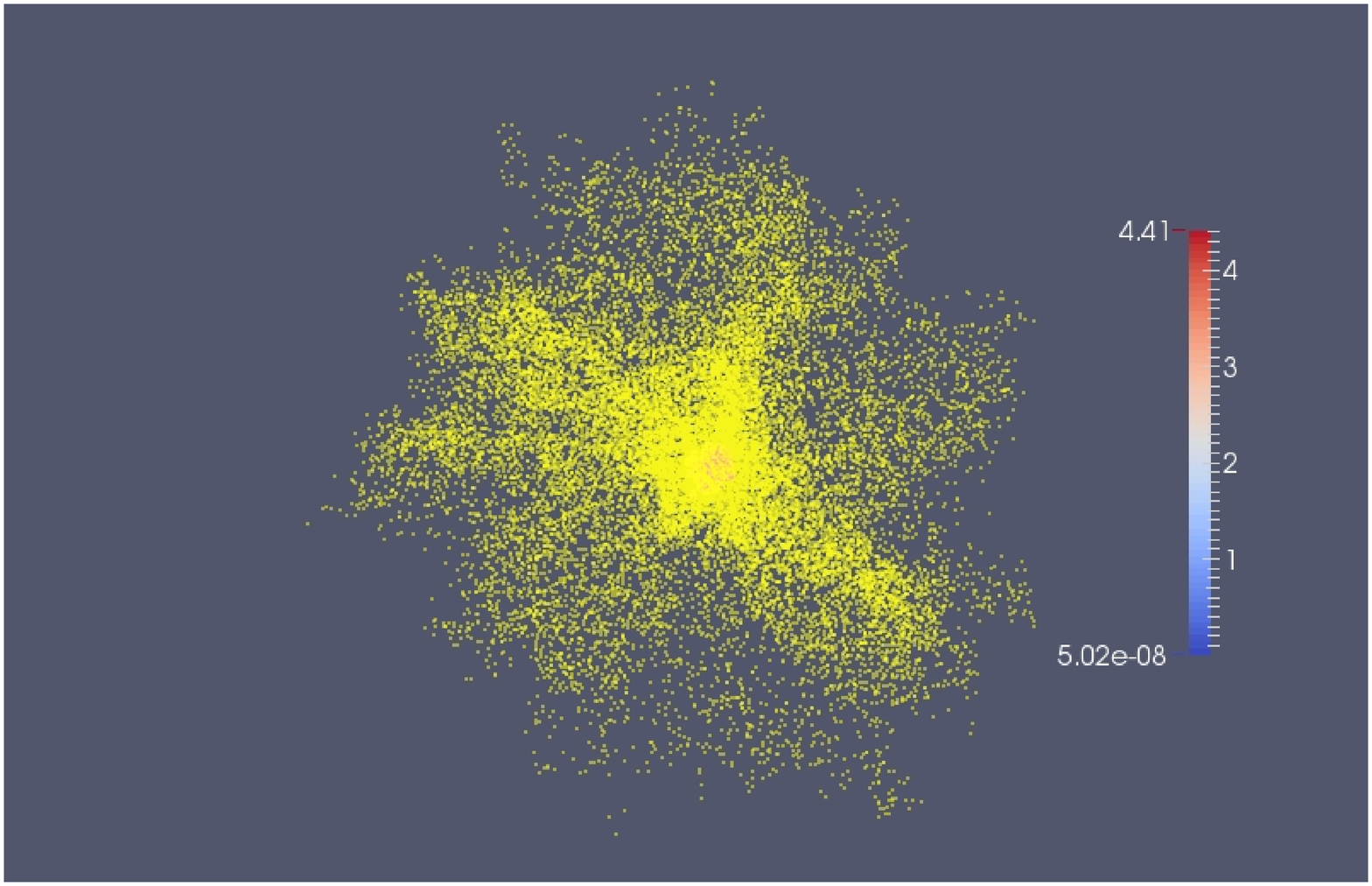} \\
\includegraphics[width=2.0 in]{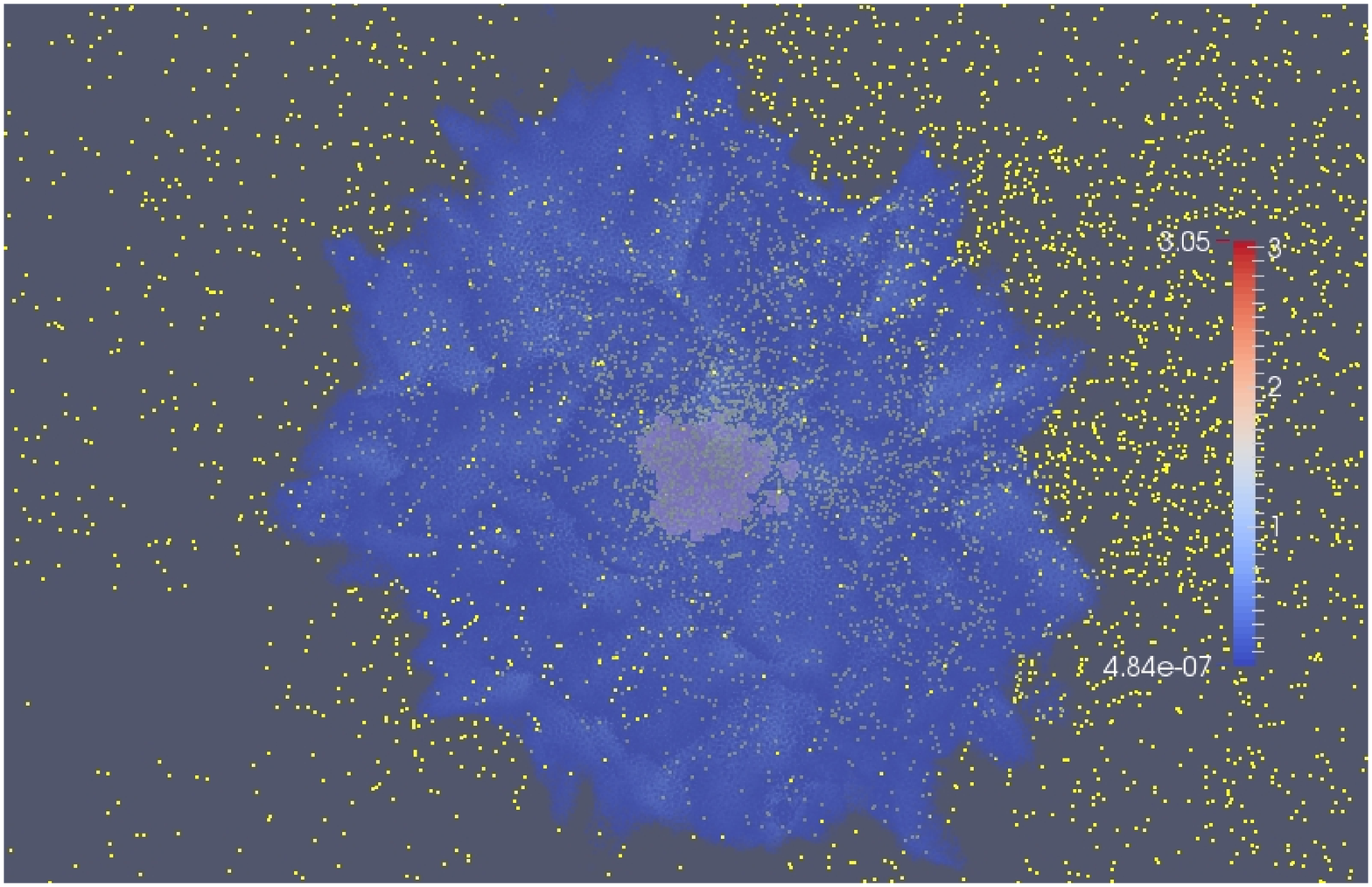} & \includegraphics[width=2.0 in]{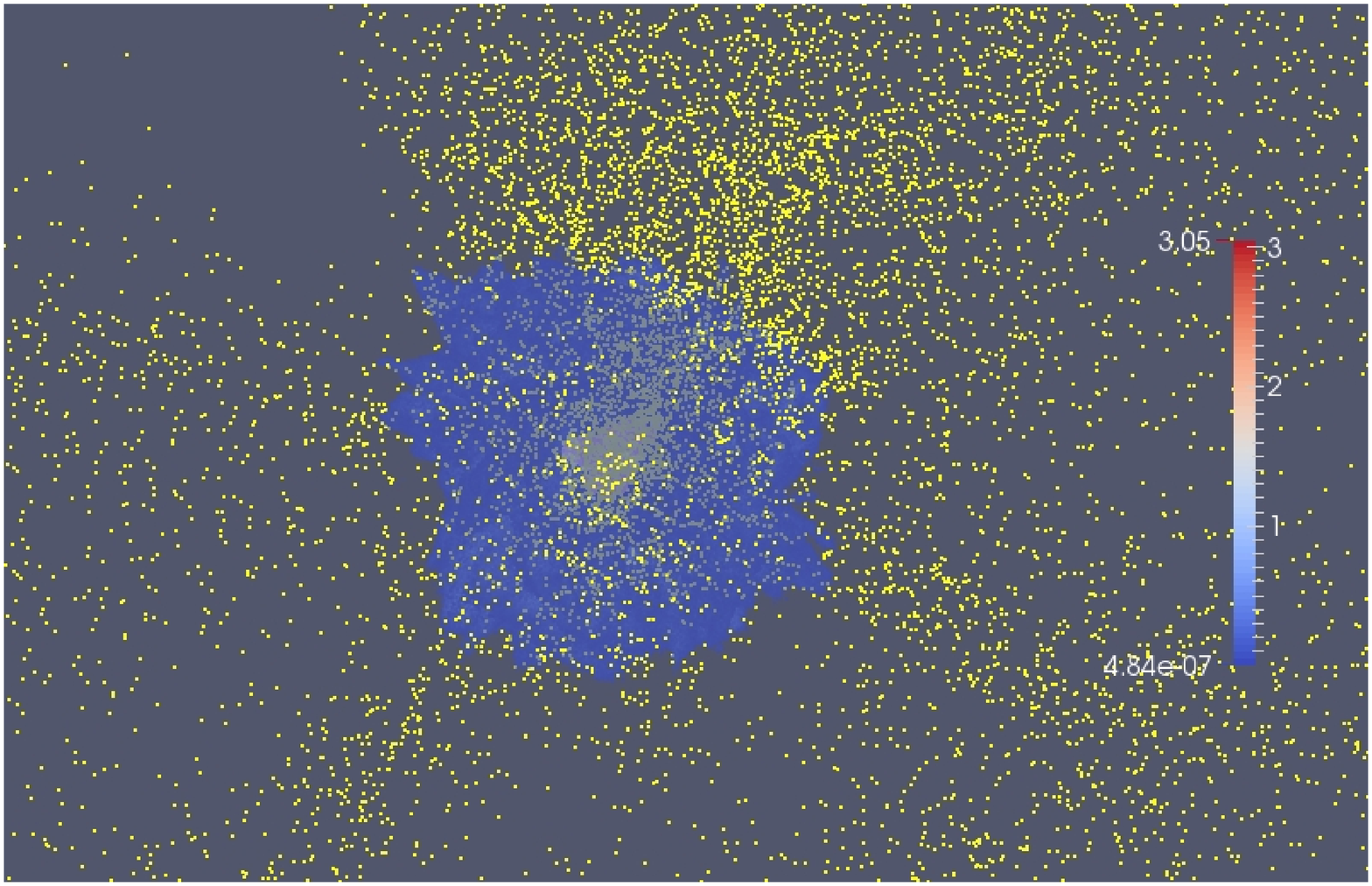} \\
\end{tabular}
\caption{ \label{vis3dGV2} $3D$ plot of the turbulent clump (in blue) with winds (in
yellow) for models (top) $Tur+Win$ and (bottom) $Tur+Win+Col$,
corresponding to last snapshot available.}
\end{center}
\end{figure}

In Fig.\ref{vis3dN} we present $3d$ plots of the densest particles for
both models. In Figs.\ref{vis3dGV} and \ref{vis3dGV2} we present
$3d$ plots at two different times in which one can distinguish the wind and
the gas particles.

We emphasize that the fraction of gas swept out by the winds is
really significant: hundreds of solar masses move even far beyond
the clump radius $R_0$. In the fourth column of
Table~\ref{tab:models}, we show that the initial mass contained
within the radius $R_s$( which defines the outer boundary of the
initial wind configuration) is around 7 and 5 $M_{\odot}$,
respectively. It is then surprising to notice that the total mass
dragged outside $R_0$ by the end of the simulation time is around
359 and 422 $M_{\odot}$, respectively, as is shown in the fifth
column of Table~\ref{tab:models}.

Lastly, in columns 6 and 7 of Table~\ref{tab:models}, we show
the velocities attained by those particles located far outside
the initial clump. Furthermore, we mention that these velocities are
not terminal, but are only the velocities during the time we follow
these simulations.

\section{Discussion}
\label{sec:dis}

We mentioned in Sect. \ref{subs:energies} that the clump
under consideration here is initially given a ratios of thermal ($\alpha$ ) and
kinetic energies ($\beta$) to the gravitational energy,
respectively, such that the clump collapse is greatly
favored. In fact, as we see in Sect.\ref{ssec:evoturbclump}, the clump presents
a strong collapse towards the central region, in which no
fragmentation is observed. This behavior can be seen 
in other turbulent simulations, when the turbulent Jeans mass ( in analogy to 
the thermal Jeans mass) is large, so that only one turbulent Jeans 
mass is "contained" in the total clump mass, see Ref.\cite{palau}.

\begin{figure}
\begin{center}
\begin{tabular}{cc}
\includegraphics[width=2.0 in]{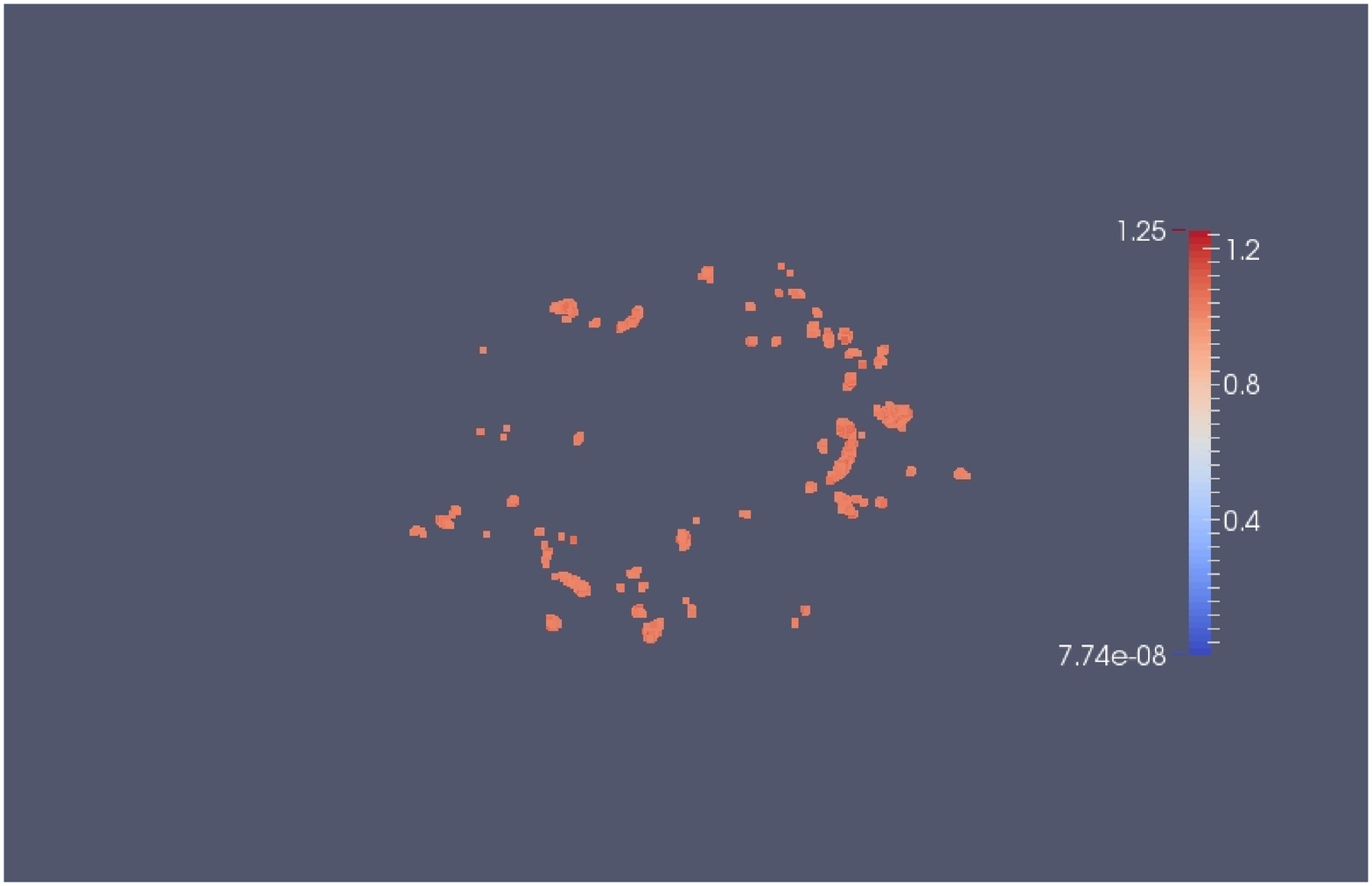} & \includegraphics[width=2.0 in]{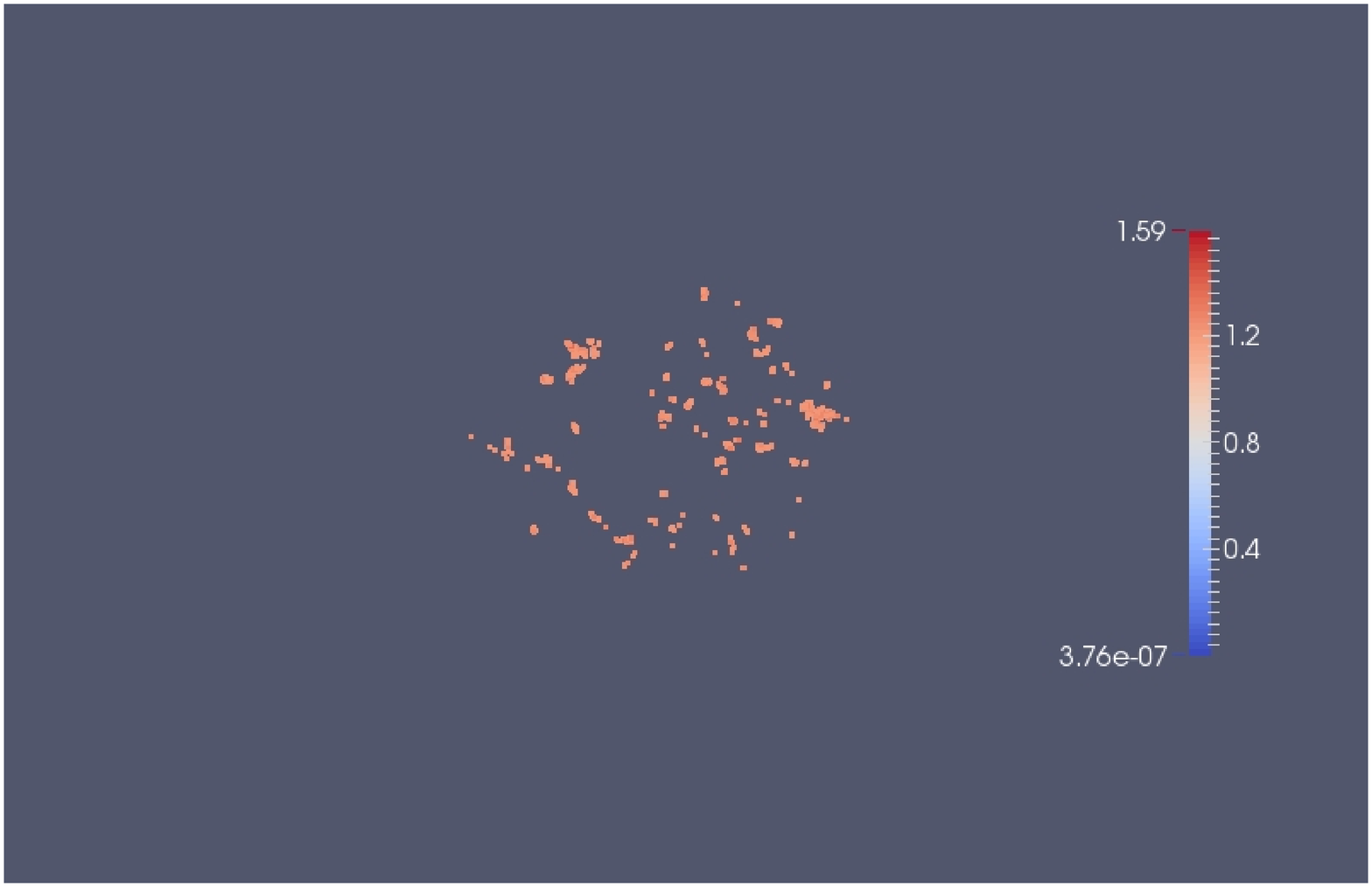} \\
\end{tabular}
\caption{ \label{vis3dN} $3D$ plot of the densest particles for models
(left) $Tur+Win$ and (right) $Tur+Win+Col$
at the corresponding snapshots shown in the second lines of Figs. \ref{NSTVRR6pp} and
\ref{NSTVRRJ7pp}, respectively.}
\end{center}
\end{figure}

The winds are activated in a very early stage of the clump evolution, when turbulence is
still ongoing. Soon after, turbulence quickly decays, so that a purely gravity driven
collapse will take place in the clump. As we see in Sect.\ref{ssec:effewindsturbclump}, the
winds act firstly as a disruptive perturbation on the clump's near environment.
Despite of this, the clump particles quickly reform this evacuated
central region. Now, in Ref.\cite{AdvAst} we found that the wind models
show a strong tendency to form accretion centers in the central region of the
clump. It was noticed there that great differences appeared when we compared the number and
location of the accretion centers obtained for the turbulent clump and for
the clump in the presence of wind. Unfortunately, these results are not
included in this paper for a lack of space. Although we try to indicate them by
means of the first panel of Fig.\ref{vis3dN}.

Thus, in view of these results, we consider that the expected fate of the
turbulent clump is really affected by the wind-clump interaction, as compared with the
purely gravity driven collapse, despite that it seems to be the dominant
physics in determining the time evolution of the clump. This behavior appears to be the case in
general, at least for observed gas structures around 0.1 pc in size, as
is discussed in Ref.\cite{palau}.

It must be noted that in this paper, as a first approximation, the
wind emission is a unique event, as it is emitted just for only
one time. It would be interesting to make that the winds be created
and emitted in each time step, so that a continuous emission rate
can be simulated. In this case, one would expect that the collapse
delay be larger and that the disruption wind effects be more
important too.

\newpage 
\section{Concluding Remarks}
\label{sec:conclu}

A star formation scenario based only on the collapse of turbulence
gas structures gives a very highly efficient transformation of the
gas into protostars.  This points is in contradiction with
observations. Besides, a physical system showing simultaneously
in-fall and outflow motions are observed in cluster like NGC 1333
and NGC 2264, see Refs.\cite{walsh} and \cite{peretto},

Because of this, another scenario must be considered, or at least a
theoretical complement to the turbulent model is needed. As we have
shown in this paper, the winds must be considered as an additional
ingredient to complement the turbulent model with the hope that this
new model can alleviate some of the problems mentioned above as they
make a delay in the runaway collapse of the clump, see
Fig.~\ref{velturviento}.

We have shown here that all the wind models show a strong tendency to 
form accretion centers in the central region of the clump. It must 
be noted that some differences appear when we compare the accretion 
centers obtained for the wind models, as can be seen in 
Fig.~\ref{vis3dN}, ~\ref{vis3dGV} and ~\ref{vis3dGV2}.
\subsubsection*{Acknowledgments.}{We would like to thank ACARUS-UNISON
for the use of their computing facilities in the preparation of this paper.}



\end{document}